\documentclass{WileyMSP-template}

\usepackage{textcomp}
\usepackage{upgreek}
\usepackage{tabularx, booktabs}
\usepackage{amssymb}
\usepackage{parskip}
\usepackage{hyperref}
\usepackage[labelfont=bf]{caption}
\usepackage{appendix}
\usepackage{booktabs}

\begin{document}
\raggedright
\pagestyle{plain}

\title{Beta Tantalum Transmon Qubits with Quality Factors Approaching 10 Million}

\maketitle

\author{Atharv Joshi}
\author{Apoorv Jindal}
\author{Paal H. Prestegaard}
\author{Faranak Bahrami}
\author{Elizabeth Hedrick}
\author{Matthew P. Bland}
\author{Tunmay Gerg}
\author{Guangming Cheng}
\author{Nan Yao}
\author{Robert J. Cava}
\author{Andrew A. Houck}
\author{Nathalie P. de Leon*}

\begin{affiliations}
Atharv Joshi, Dr. Apoorv Jindal, Paal H. Prestegaard, Dr. Faranak Bahrami, Elizabeth Hedrick, \\
Matthew P. Bland, Tunmay Gerg\footnote{Present address: Yale University, Departments of Applied Physics and Physics, New Haven, CT 06511, USA.}, Prof. Andrew A. Houck, Prof. Nathalie P. de Leon \\
Princeton University, Department of Electrical and Computer Engineering, Princeton, NJ 08544, USA\\
Email Address: npdeleon@princeton.edu \\

\medskip

Guangming Cheng, Nan Yao \\
Princeton University, Princeton Materials Institute, Princeton, NJ 08544, USA \\

\medskip

Robert J. Cava \\
Princeton University, Department of Chemistry, Princeton, NJ 08544, USA \\

\end{affiliations}


\justifying

\begin{abstract}

Tantalum-based transmon qubits are a promising platform for building large-scale quantum processors. So far, these qubits have been made from tantalum films grown exclusively in the alpha phase ($\alpha$-Ta). The beta phase of tantalum ($\beta$-Ta) readily nucleates at room temperature, making it attractive for scalable qubit fabrication. However, $\beta$-Ta is widely believed to be detrimental to qubit performance because it has a lower superconducting critical temperature than $\alpha$-Ta. We challenge this prevailing belief by fabricating low-loss transmon qubits from $\beta$-Ta films on sapphire. Across 11 qubits, the mean time-averaged quality factor is $(5.6 \pm 2.3) \times 10^6$, with the best qubit recording a time-averaged quality factor of $(10.1 \pm 1.3) \times 10^6$. Resonator studies demonstrate that the dominant microwave loss channel is surface two-level systems, with the surface loss contribution for $\beta$-Ta being about twice that of $\alpha$-Ta. $\beta$-Ta films exhibit significant kinetic inductance, consistent with an estimated magnetic penetration depth of $(1.78 \pm 0.02) \ \upmu$m. This work establishes $\beta$-Ta on sapphire as a material platform for realizing low-loss transmon qubits and other superconducting devices such as compact resonators, kinetic inductance detectors, and quasiparticle traps.

\end{abstract}

\clearpage
\setlength{\parindent}{12pt}

Superconducting qubits are a leading platform for building large-scale quantum processors, having achieved early milestones in quantum error correction \cite{sivak2023realtime, ni2023beating, google2025quantum, putterman2025hardware} and quantum simulation of many-body systems \cite{kim2023evidence, andersen2025thermalization, google2025observation, liu2026prethermalization}. Advancing the scale and performance of superconducting quantum processors requires understanding and eliminating material-related microwave losses in the constituent qubits \cite{deleon2025how}. Systematic improvements in material selection and device fabrication have recently led to sustained gains in the lifetimes and coherence times of superconducting qubits. Lifetimes exceeding 100 $\upmu$s have been realized in several material platforms \cite{bland2025millisecond, deng2023titanium, biznarova2024mitigation, tuokkola2025methods}, with tantalum (Ta)-based 2D transmon qubits \cite{bland2025millisecond} and quantum memories \cite{ganjam2024surpassing} achieving millisecond-scale coherence. 

State-of-the-art Ta superconducting qubits have thus far been made exclusively from films sputtered in the body-centered cubic (b.c.c.) phase ($\alpha$-Ta) \cite{bland2025millisecond, place2021new, wang2022towards, olszewski2026krypton, bu2025room}. While Ta thin films nucleate more readily at room temperature in the metastable tetragonal phase ($\beta$-Ta) \cite{read1965new}, $\beta$-Ta is considered to be detrimental to qubit performance because it has a lower superconducting critical temperature ($T_\mathrm{c}$) than $\alpha$-Ta (Table \ref{table:properties}). Superconducting qubits typically operate at few-photon powers and millikelvin temperatures. At these temperatures, we expect negligible microwave losses due to quasiparticles (QP) in thermal equilibrium, even with the low $T_\mathrm{c}$ of $\beta$-Ta. Assuming an equilibrium temperature of 20 mK, BCS theory predicts the fraction of Cooper pairs dissociated into QPs in a superconductor with $T_\mathrm{c} = 0.5$ K to be less than $10^{-19}$ \cite{glazman2021bogoliubov}. In practice, the observed fraction of dissociated Cooper pairs is several orders of magnitude larger ($10^{-9}-10^{-5}$), which implies that QP-induced losses in superconducting qubits primarily arise from non-equilibrium QPs \cite{glazman2021bogoliubov}. However, in the qubit operating regime, microwave losses in many materials, including $\alpha$-Ta, are dominated by defects known as two-level systems (TLS), and not by QPs \cite{gao2008physics, muller2019towards, crowley2023disentangling}.

Experimental reports comparing microwave losses in $\alpha$-Ta and $\beta$-Ta are scarce and inconsistent. Measurements on $\alpha$-Ta resonators reveal TLS loss contributions from both the surface and bulk dielectrics \cite{crowley2023disentangling}. Surface TLSs primarily reside in the native oxide, which predominantly consists of tantalum pentoxide ($\mathrm{Ta_2O_5}$) \cite{mclellan2023chemical, oh2024structure}. The native oxide of $\beta$-Ta is similarly composed of $\mathrm{Ta_2O_5}$ \cite{whitacre2001real}. Despite this similarity, three studies find an order-of-magnitude higher loss with weak power dependence in $\beta$-Ta resonators, indicating a non-TLS origin \cite{singer2024tantalum, urade2024microwave, arlt2025high}. In contrast, two studies find comparable power-dependent losses in the presence of minority \cite{dhundhwal2025high} or majority \cite{mcfadden2025interface} $\beta$ phase in mixed $\alpha$-$\beta$-Ta resonators. A natural question is whether the microwave loss channels in $\beta$-Ta superconducting devices differ from those in $\alpha$-Ta.

Here, we fabricate superconducting resonators and transmon qubits from $\beta$-Ta films deposited on sapphire at room temperature. We measure shifts in resonator frequencies that are consistent with a large kinetic inductance contribution, and we quantify the sheet kinetic inductance ($L_\mathrm{k/\square}$) of the films to ensure reliable device design. We study the power and temperature dependence of the resonator internal quality factor ($Q_\mathrm{int}$), and show that microwave losses at few-photon powers and millikelvin temperatures are dominated by TLSs. We find that the TLS loss scales linearly with the resonator surface participation ratio (SPR) \cite{wang2015surface}, indicating that the dominant loss channel is surface TLSs, and we quantify the apparent surface losses in $\beta$-Ta to be around two times higher than in $\alpha$-Ta \cite{crowley2023disentangling}. Transmon qubits fabricated from $\beta$-Ta films exhibit relaxation times ($T_\mathrm{1}$) that are in line with the loss limit imposed by surface TLSs. The best qubit, with a frequency ($f_\mathrm{q}$) of $2.74$ GHz, has a time-averaged relaxation time ($\overline{T}_\mathrm{1}$) of $(585 \pm 75) \ \upmu$s, which corresponds to a time-averaged quality factor ($\overline{Q} = 2\pi f_\mathrm{q} \overline{T}_{\mathrm{1}}$) of $(10.1 \pm 1.3) \times 10^6$, over 85 hours of continuous measurement. The mean time-averaged quality factor across 11 qubits is $(5.6 \pm 2.3) \times 10^6$, a performance that is comparable to that of the best aluminum (Al) \cite{biznarova2024mitigation}, niobium (Nb) \cite{tuokkola2025methods, bal2024systematic, gordon2022environmental, kono2024mechanically}, and $\alpha$-Ta on sapphire \cite{wang2022towards} transmons. Our results establish $\beta$-Ta on sapphire as a material platform for realizing low-loss transmon qubits.

\begin{table}[b]
    \caption{Salient material properties of sputtered Ta \cite{read1965new, baker1972preparation, bahrami2026vortex, schwartz1972temperature}.}
    \begin{tabular*}{\columnwidth}{l @{\extracolsep{\fill}} c @{\extracolsep{\fill}} c
    @{\extracolsep{\fill}} c
    @{\extracolsep{\fill}} c}
        \toprule
        Phase & Crystal structure & Lattice parameters (\AA) & $T_\mathrm{c}$ (K) & $\rho_\mathrm{n}$ ($\upmu \Omega \ \mathrm{cm}$) \\
        \midrule
        $\alpha$-Ta & b.c.c. & $a=3.3-3.4$ & $4.2-4.4$ & $1-50$ \\
        $\beta$-Ta & tetragonal & $a=5.3, \, c=9.9-10$ & $0.5-1.0$ & $150-220$\\
        \bottomrule
    \end{tabular*}
    \label{table:properties}
\end{table}

\begin{figure}[t]
\includegraphics[width=\textwidth]{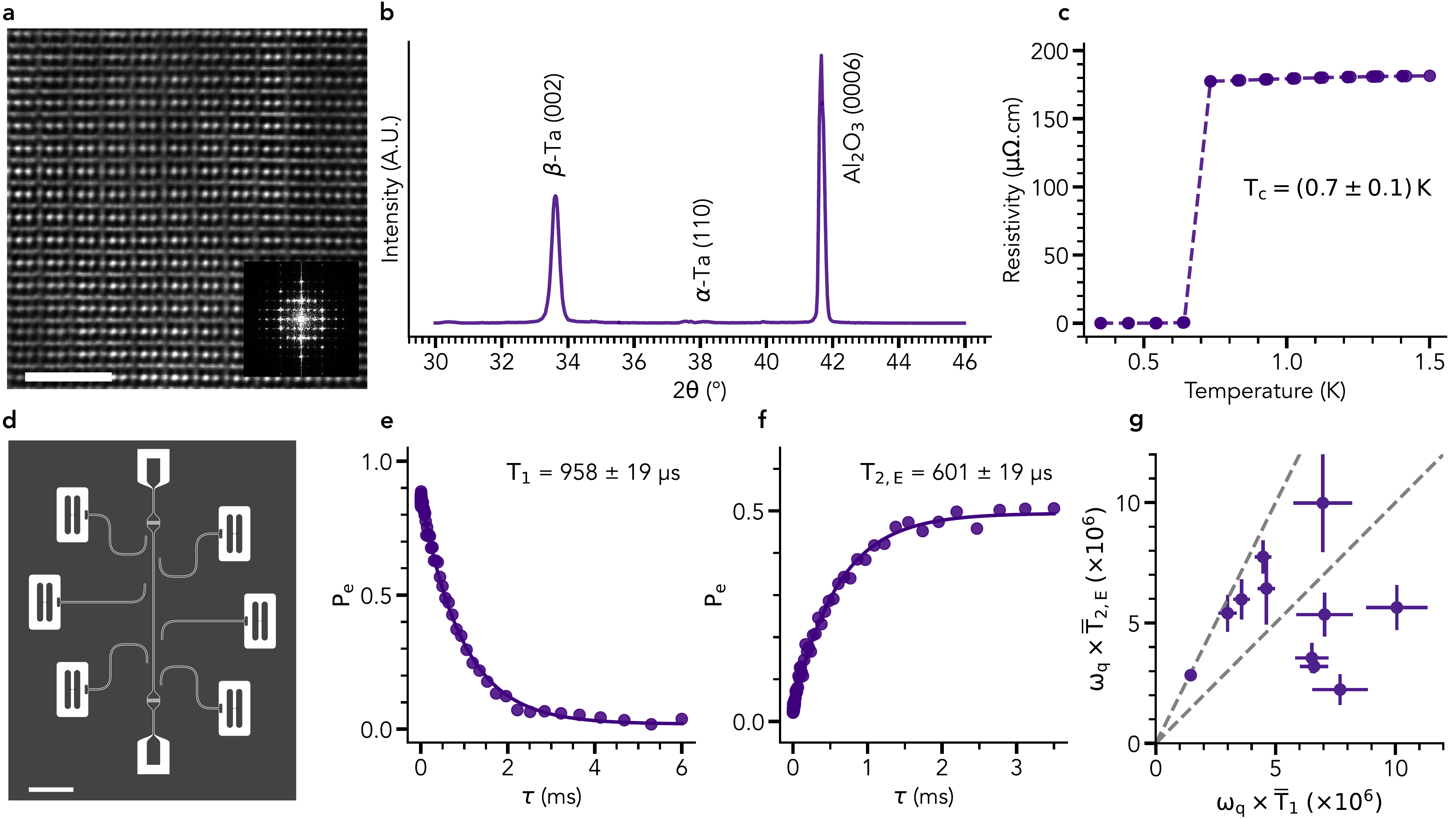}
\caption{\textbf{$\beta$-Ta transmon qubits with quality factors approaching ten million.} (a) Cross-sectional STEM image of a Ta film on sapphire. Scale bar: $2$ nm. Inset: Local area diffraction pattern identifies (001) as the preferred growth plane. (b) XRD pattern obtained from the Ta film shows intense peaks at $33.6 \degree$ and $41.7 \degree$ corresponding to $\beta$-Ta $(002)$ \cite{face1987nucleation} and $\mathrm{Al_2 O_3} \ (0006)$ respectively, along with traces of $\alpha$-Ta $(110)$. (c) Four-probe DC resistivity measurement across the superconducting transition of the Ta film. The $T_\mathrm{c}$ of $(0.7 \pm 0.1)$ K is reported as the midpoint of the temperatures of the two consecutive data points that bracket the superconducting transition, and the uncertainty in $T_\mathrm{c}$ represents the measurement temperature step size. The dashed line is a guide to the eye. (d) Schematic of a chip with 6 transmon qubits on a sapphire substrate (white), where the transmon capacitor pads and ground plane are made from $\beta$-Ta (gray). Scale bar: $1$ mm. Excited state population ($P_\mathrm{e}$) as a function of delay time ($\tau$) for a qubit with \mbox{$f_\mathrm{q} = 2.74$ GHz} in a (e) $T_\mathrm{1}$ experiment showing a maximum \mbox{$T_\mathrm{1} = (958 \pm 19) \ \upmu$s} and (f) $T_\mathrm{2, E}$ experiment showing a maximum \mbox{$T_\mathrm{2, E} = (601 \pm 19) \ \upmu$s}. Solid lines are fits to the data points and uncertainties in $T_\mathrm{1}$ and $T_\mathrm{2, E}$ represent the standard error (s.e.) estimated from the covariance matrix of the fit. (g) Plot of $\overline{T}_\mathrm{1}$ and $\overline{T}_\mathrm{2, E}$ multiplied by \mbox{$\omega_q=2 \pi f_{q}$} for $11$ qubits, where error bars represent the standard deviation (s.d.) of the $T_\mathrm{1}$ and $T_\mathrm{2, E}$ time series. The two dashed lines, in ascending order of slope, depict the \mbox{$\overline{T}_\mathrm{2, E} = \overline{T}_\mathrm{1}$} and \mbox{$\overline{T}_\mathrm{2, E} = 2 \, \overline{T}_\mathrm{1}$} limits respectively.}
\label{figure: summary}
\end{figure}

Unlike $\alpha$-Ta thin films, which require stringent growth conditions \cite{place2021new, olszewski2026krypton, bu2025room, face1987nucleation, lozano2024low, marcaud2025low, van2025cryogenic}, such as elevated temperatures or seed layers, $\beta$-Ta films can be readily grown at room temperature \cite{read1965new, baker1972preparation}. We deposit Ta films on $500 \ \upmu$m-thick c-plane sapphire substrates at room temperature by DC magnetron sputtering (Section S1, Supporting Information). Cross-sectional scanning transmission electron microscopy (STEM) reveals a tetragonal unit cell with lattice parameters \mbox{$a = 5.3$ \AA} and \mbox{$c = 10$ \AA} (Figure \ref{figure: summary}a), which is characteristic of $\beta$-Ta (Table \ref{table:properties}). X-ray diffraction (XRD) confirms the predominant crystallographic phase to be $\beta$-Ta (Figure \ref{figure: summary}b). A four-probe DC resistivity measurement as a function of temperature shows a $T_\mathrm{c}$ of \mbox{($0.7 \pm 0.1$) K} and a normal state resistivity ($\rho_\mathrm{n}$) of \mbox{($180 \pm 1) \ \upmu \Omega \ \mathrm{cm}$} (Figure \ref{figure: summary}c), which agree with the reported values for $\beta$-Ta (Table \ref{table:properties}). We pattern the $\beta$-Ta films into resonators and qubit wiring layers (Figure \ref{figure: summary}d) using photolithography and reactive ion etching (Section S2, Supporting Information). We fabricate Al/aluminum oxide/Al Josephson junctions for the qubits in an ultrahigh-vacuum electron-beam evaporation system, following procedures described in Reference \cite{bland2025millisecond}. All devices are measured in dilution refrigerators with base temperatures of $10-15$ mK (Section S3, Supporting Information). 

Despite the low $T_\mathrm{c}$ of the $\beta$-Ta films, transmon qubits with $\beta$-Ta capacitor pads show remarkably long relaxation and coherence times. The best qubit records a maximum $T_\mathrm{1}$ of $(958 \pm 19) \ \upmu$s (Figure \ref{figure: summary}e), Hahn echo coherence time ($T_\mathrm{2, E}$) of $(601 \pm 19) \ \upmu$s (Figure \ref{figure: summary}f), and Ramsey coherence time ($T_\mathrm{2, R}$) of $(408 \pm 15) \ \upmu$s (Section S4, Supporting Information). We acquire $T_\mathrm{1}$ and $T_\mathrm{2, E}$ measurements over $2-5$ days for each qubit. Across 11 qubits, the mean time-averaged relaxation time is $(264 \pm 165) \ \upmu$s and the mean time-averaged Hahn echo coherence time is $(214 \pm 80) \ \upmu$s, with 4 qubits attaining nearly lifetime-limited coherence, \mbox{$\overline{T}_\mathrm{2, E} \sim 2 \, \overline{T}_\mathrm{1}$} (Figure \ref{figure: summary}g). 

\begin{figure}[t]
\includegraphics[width=\textwidth]{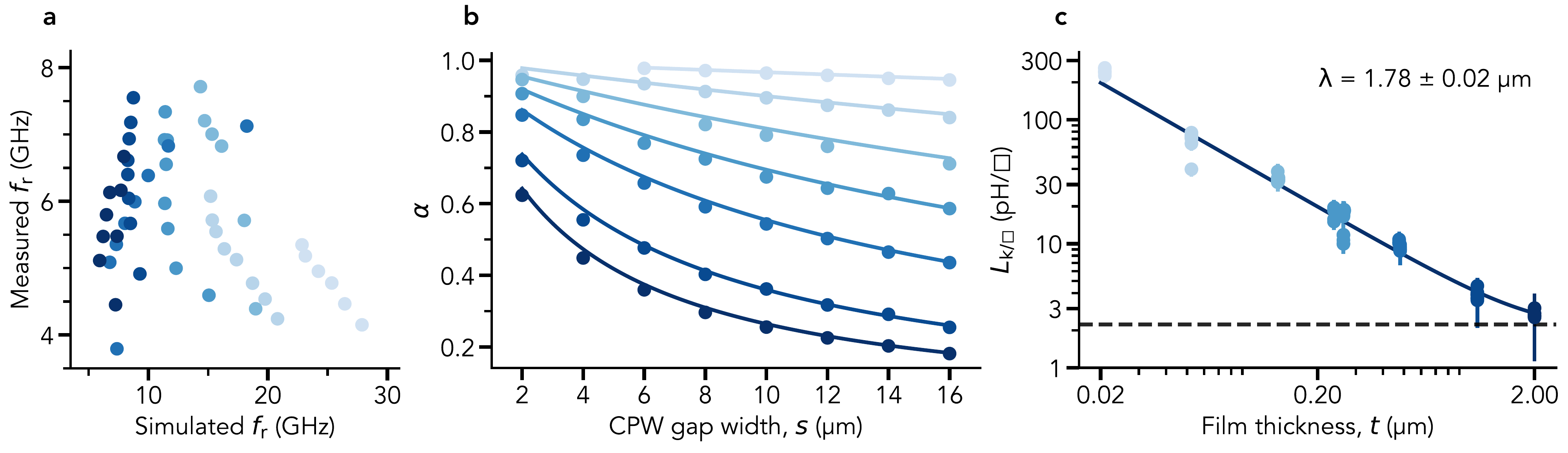}
\caption{\textbf{Quantifying the kinetic inductance of $\beta$-Ta films}. (a) Scatter plot of measured and simulated resonance frequencies ($f_\mathrm{r}$) of 80 CPW resonators fabricated from 11 $\beta$-Ta films with different thicknesses. Film thickness ($t$) ranges from $0.02 \ \upmu$m (lightest blue) to $2.00 \ \upmu$m (darkest blue) and roughly doubles with each intermediate shade of blue. The measured $f_\mathrm{r}$ is always lower than the simulated $f_\mathrm{r}$, with the largest shifts to lower frequencies occurring in the thinnest films. (b) Plot of the kinetic inductance fraction ($\alpha$) versus CPW gap width ($s$) for a subset of 7 films. We use electromagnetic simulations (Section S7, Supporting Information) to obtain an expected $\alpha$ for each $s$ (solid lines). (c) The sheet kinetic inductance ($L_\mathrm{k/\square}$) extracted from each resonator, as a function of $t$, lies on a curve that is well-described by a single magnetic penetration depth \mbox{$\lambda = 1.78 \, \pm \, 0.02 \ \upmu$m} (solid line). The uncertainty in $\lambda$ represents the s.e. estimated from the covariance matrix of the fit. The $L_\mathrm{k/\square}$ saturates to $2.2 \ \mathrm{pH/\square}$ in the bulk limit (dashed line).}
\label{figure: ki}
\end{figure}

To investigate the material loss channels limiting qubit performance, we measure microwave losses in coplanar waveguide (CPW) and lumped-element (LE) resonators made from $\beta$-Ta films (Section S5, Supporting Information), following the methodology developed in Reference \cite{crowley2023disentangling}. We measure the resonance frequencies ($f_\mathrm{r}$) of the resonators to be lower than their simulated $f_\mathrm{r}$, with CPW resonators displaying particularly large shifts to lower frequencies (Figure \ref{figure: ki}a). The measured $f_\mathrm{r}$ is extracted from a transmission measurement (Section S6, Supporting Information) and the simulated $f_\mathrm{r}$ is obtained from electromagnetic simulations of the resonator geometry (Section S7, Supporting Information). Modeling this shift to lower frequencies is crucial for identifying resonators during loss measurements and designing qubit wiring layers. 

We hypothesize that this shift originates from the kinetic inductance of $\beta$-Ta films, which has been reported in previous works \cite{kouwenhoven2023resolving, de2025recombination}. We validate this hypothesis by fabricating 80 CPW resonators with varying gap widths ($s=2-16 \ \upmu$m) from 11 films with different thicknesses ($t=0.02-2.00 \ \upmu$m), as kinetic inductance has a distinct dependence on $s$ and $t$ \cite{gao2008physics}. The frequency shift is quantified by $\alpha = 1 - (f_\mathrm{r, \, measured} / f_\mathrm{r, \, simulated})^2$, which we interpret as the effective kinetic inductance fraction. Across all CPW resonators, $\alpha$ increases systematically with decreasing $s$ and $t$ (Figure \ref{figure: ki}b). We model this behavior of $\alpha$ using a geometric inductance, $L_\mathrm{g}(s)$, and a kinetic inductance, $L_\mathrm{k}(s, t)$, acting in series. We use electromagnetic simulations of the resonator surface current distributions to determine $L_\mathrm{g}(s)$ and map the $L_\mathrm{k}(s, t)$ of the resonator to the sheet kinetic inductance of the film, $L_\mathrm{k/\square}(t)$ (Section S7, Supporting Information). Resonators fabricated on films with the same thickness yield comparable $L_\mathrm{k/\square}(t)$ values (Figure \ref{figure: ki}c), and the data are well-described by a single curve given by \cite{klein1990effective, lopeznunez2025superconducting}

\begin{equation} \label{eqn: lksq}
    L_\mathrm{k/\square}(t) = \mu_0 \lambda \coth\Big(\frac{t}{\lambda}\Big),
\end{equation}

\noindent where $\mu_0$ is the vacuum magnetic permeability, $t$ is measured using atomic force microscopy, and $\lambda$ is the magnetic penetration depth. The fitted $\lambda$ of $(1.78 \pm 0.02) \ \upmu$m is around $0.1 \ \upmu$m larger than an estimate calculated assuming the BCS dirty limit (Section S7, Supporting Information). Our approach reliably estimates $L_\mathrm{k/\square}$ for thin films ($t < \lambda$) typically used for fabricating superconducting devices. For instance, we expect $L_\mathrm{k/\square} \sim 10-100\, \mathrm{pH/\square}$ for $t \sim 400 - 40$ nm.  

\begin{figure}[t]
\includegraphics[width=\textwidth]{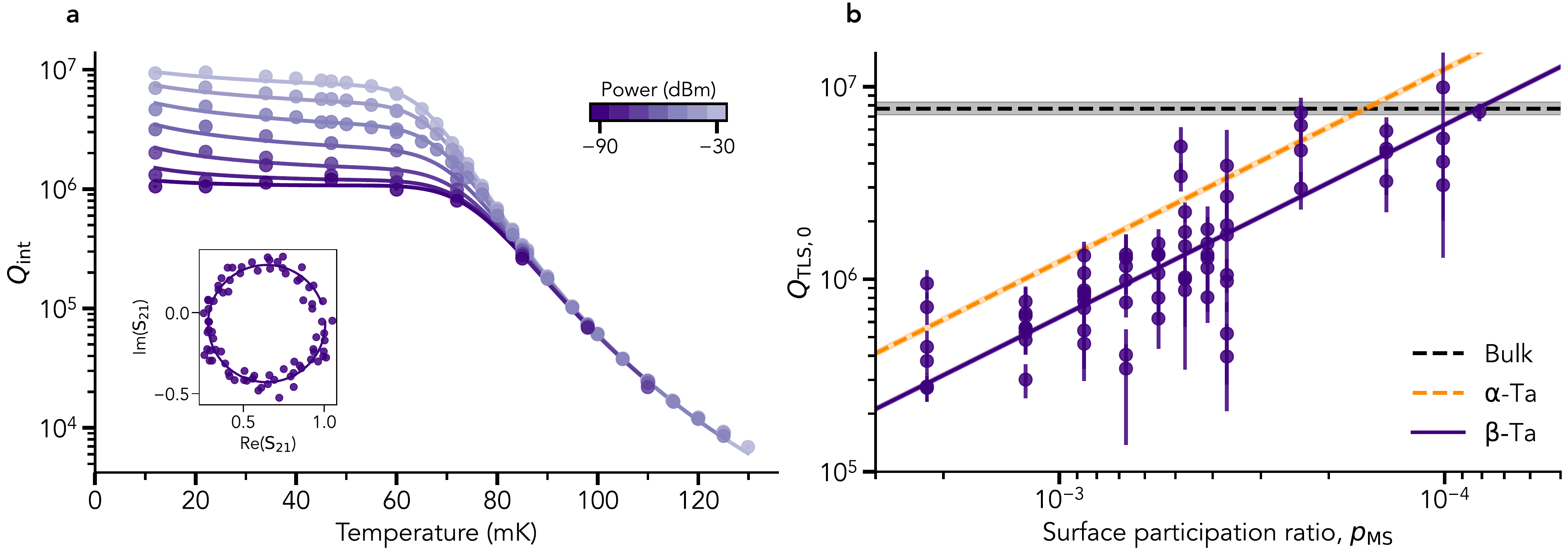}
\caption{\textbf{Quantifying microwave losses in $\beta$-Ta resonators.} (a) $Q_\mathrm{int}$ as a function of microwave probe power and mixing chamber stage temperature for a CPW resonator (\mbox{$s = 6 \ \upmu$m},  \mbox{$t = 263$ nm}). Solid lines are fits to a model that parametrizes losses due to TLSs, QPs, and other power- and temperature-independent channels (Section S8, Supporting Information). Inset: Real and imaginary components of the transmission coefficient ($S_\mathrm{21}$) at a power of $-90$ dBm and a temperature of $11$ mK, acquired using a homophasal point distribution \cite{baity2024circle}, and fit to a circle \cite{probst2015efficient} (solid line) to extract $Q_\mathrm{int}$ (Section S6, Supporting Information). Error bars of $Q_\mathrm{int}$ represent the s.e. estimated from the covariance matrix of the circle fit, and are smaller than the data points. (b) $Q_\mathrm{TLS, 0}$ extracted from power-temperature sweeps of $Q_\mathrm{int}$ for 71 $\beta$-Ta resonators scales inversely with $p_\mathrm{MS}$. Error bars of $Q_\mathrm{TLS, 0}$ represent the s.e. estimated from the covariance matrix of the fit. We choose $p_\mathrm{MS}$ for consistency with previous literature \cite{bland2025millisecond, crowley2023disentangling, wang2015surface}. The data are well-described by a single surface loss tangent (solid line). The loss tangents corresponding to the $\alpha$-Ta surface (dashed orange line) and the bulk substrate (dashed black line) and their uncertainties (shaded regions) are taken from Reference \cite{crowley2023disentangling} for comparison.}
\label{figure: tls}
\end{figure}

Superconductors with micron-scale $\lambda$ values often exhibit increased inductive losses, which may decrease resonator $Q_{\mathrm{int}}$ \cite{charpentier2025universal}. We extract the $Q_{\mathrm{int}}$ of $\beta$-Ta resonators from a transmission measurement (Figure \ref{figure: tls}a inset; Section S6, Supporting Information), and find that a high $L_\mathrm{k/\square}$ coexists with high $Q_{\mathrm{int}}$. For instance, a CPW resonator ($s = 8 \ \upmu$m, $t = 21$ nm) fabricated from a film with an $L_\mathrm{k/\square}$ of $(259 \pm 33) \ \mathrm{pH/\square}$ exhibits a $Q_{\mathrm{int}}$ of $(1.6 \pm 0.1) \times 10^6$ at low powers and temperatures (Section S5, Supporting Information); this is among the highest combinations of $L_\mathrm{k/\square}$ and $Q_{\mathrm{int}}$ reported in the literature \cite{charpentier2025universal}.

We measure $Q_{\mathrm{int}}$ at various powers ($\overline{n}$) and temperatures ($T$) to identify the dominant loss channels in $\beta$-Ta resonators. Power is expressed as an effective intracavity photon number $\overline{n}$ after accounting for line attenuation in the measurement chain. A representative CPW resonator (Figure \ref{figure: tls}a) shows $Q_{\mathrm{int}}$ ranging from $10^4 - 10^7$. At $T < 50$ mK,  $Q_{\mathrm{int}}$ decreases with decreasing power; this behavior is consistent with losses due to saturable TLSs \cite{gao2008physics, muller2019towards}. Above 70 mK, $Q_{\mathrm{int}}$ decreases exponentially with temperature due to the generation of thermal QPs. We fit $Q_\mathrm{int}(\overline{n}, \ T)$ to a model \cite{crowley2023disentangling} that separates losses due to TLSs, thermal QPs, and other power- and temperature-independent channels (Section S8, Supporting Information). At the lowest measured power and temperature, TLS losses account for about $98 \%$ of total internal losses in the device in Figure \ref{figure: tls}a. We quantify the TLS loss contribution by $Q_{\mathrm{TLS, 0}}$, a material-dependent fit parameter that constitutes the inverse linear absorption due to TLSs. 

We quantify the relative contributions of the surface and bulk TLS baths by studying the dependence of $Q_{\mathrm{TLS, 0}}$ on the resonator SPR. In the CPW resonators, the SPR of the metal-substrate interface ($p_\mathrm{MS}$) ranges from $(0.4-2) \times 10^{-3}$, and the LE resonators extend this range to include $p_\mathrm{MS}$ down to $8 \times 10^{-5}$ (Section S5, Supporting Information). All resonators undergo a post-fabrication buffered oxide etch treatment to reduce the native oxide thickness and remove hydrocarbons associated with fabrication contamination \cite{crowley2023disentangling, mclellan2023chemical}. The $Q_{\mathrm{TLS, 0}}$ of 71 CPW and LE resonators increases linearly with decreasing $p_\mathrm{MS}$ (Figure \ref{figure: tls}b). The linear dependence indicates that the losses are dominated by surface TLSs across this range of $p_\mathrm{MS}$, and we do not observe any contribution from bulk TLSs in this range. We extract an effective surface loss tangent ($\mathrm{tan} \ \delta$) of $(1.6 \pm 0.1) \times 10^{-3}$, which is about twice that of $\alpha$-Ta reported in Reference \cite{crowley2023disentangling}. The data are well-explained by a single $\mathrm{tan} \ \delta$ regardless of metal film thickness (Section S9, Supporting Information).

We explore potential sources of the higher apparent surface TLS loss compared to $\alpha$-Ta in Figure \ref{figure: tls}b by examining the metal-air (MA) and MS interfaces of the resonators. As the native oxide at the MA interface is a major source of TLSs, we study its composition using X-ray photoelectron spectroscopy (XPS). We fit XPS scans of the Ta4f peaks to extract an effective oxide thickness ($t_\mathrm{TaOx}$) of $(2.9 \pm 0.1)$ nm, which is consistent with cross-sectional STEM images of the MA interface (Section S9, Supporting Information). This value of $t_\mathrm{TaOx}$ is about $1.2$ times greater than that for $\alpha$-Ta reported in Reference  \cite{mclellan2023chemical}. The increased $t_\mathrm{TaOx}$ can account for $20\%$ of the higher surface losses in $\beta$-Ta within a model where MA interface loss scales linearly with the oxide thickness \cite{crowley2023disentangling} (Section S9, Supporting Information). Such a model assumes a laterally uniform oxide layer. However, scanning electron microscopy images of device sidewalls show protrusions with lateral dimensions of about $100$ nm that contribute to increased sidewall roughness (Section S9, Supporting Information). These features can increase the effective oxide volume at the MA interface, and may lead to enhanced surface losses when present in regions with high electric field participation,  such as sidewalls \cite{chang2025eliminating}. We also acquire cross-sectional STEM images of the MS interface and find a roughly 5 nm-thick amorphous $\beta$-Ta layer at the epitaxial interface (Section S9, Supporting Information). This amorphous layer is another potential source of TLS loss, and its loss contribution may be quantified in future work by measuring devices that are designed to independently vary the participation ratios at the MS, MA, and substrate-air interfaces \cite{woods2019determining, zemlicka2023compact, lei2023characterization}.

\begin{figure}[t]
\includegraphics[width=\textwidth]{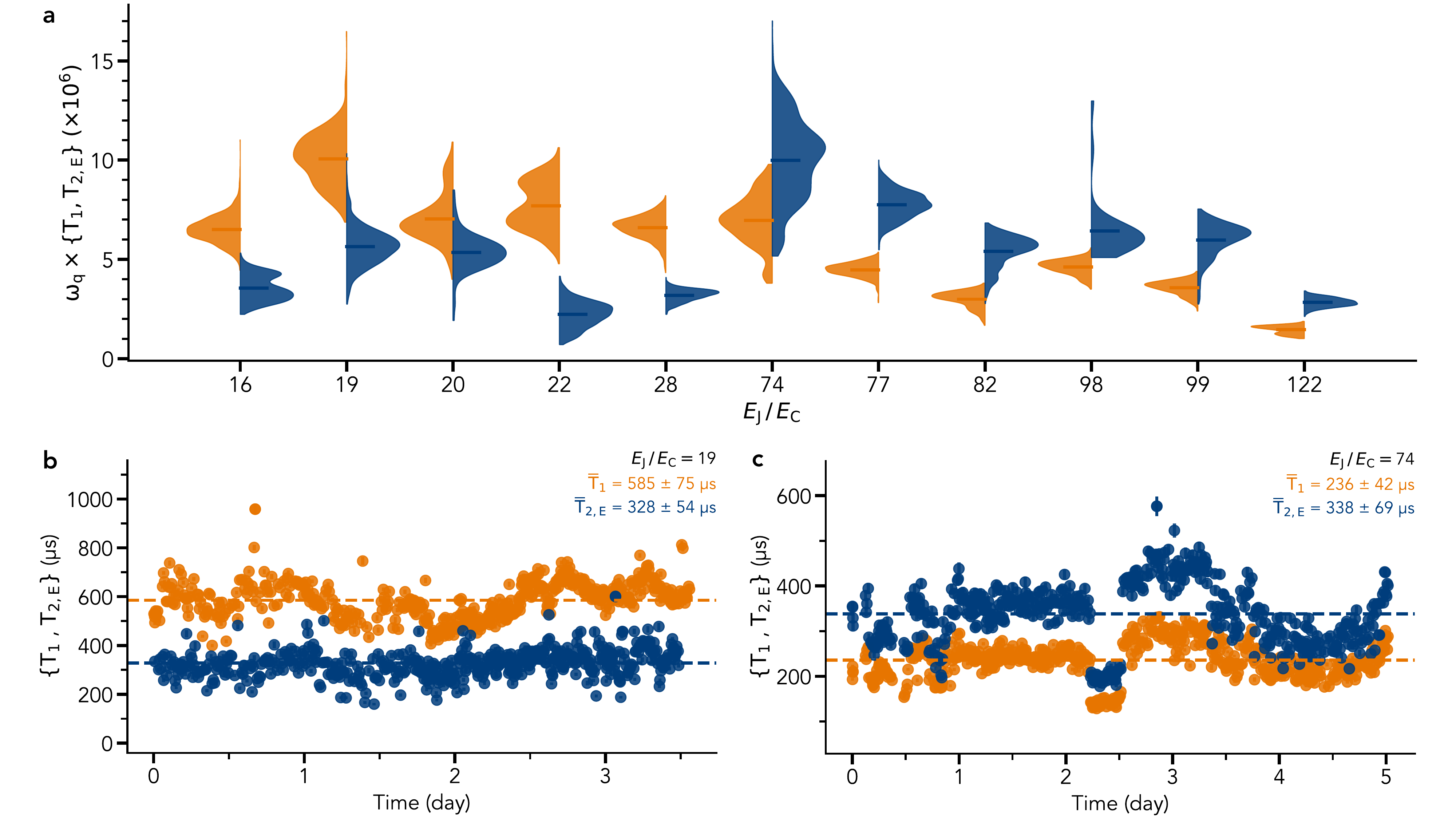}
\caption{\textbf{Performance of $\beta$-Ta transmon qubits.} (a) Split violin plot showing the distribution of all $T_\mathrm{1}$ (left, orange) and $T_\mathrm{2, E}$ (right, blue) measurement results multiplied by $\omega_\mathrm{q}$ for all $11$ qubits ordered by increasing \mbox{$E_\mathrm{J}/E_\mathrm{C}$}. The $\overline{T}_\mathrm{1}$ and $\overline{T}_\mathrm{2, E}$ values multiplied by $\omega_\mathrm{q}$ are depicted by horizontal lines within each distribution. The distributions are generated by applying Gaussian kernel density estimation to the sampled data. Results of successive $T_\mathrm{1}$ (orange) and $T_\mathrm{2, E}$ (blue) measurements over a few days acquired on (b) an OCS transmon \mbox{($f_\mathrm{q} = 2.74$ GHz)} with \mbox{$E_\mathrm{J}/E_\mathrm{C} = 19$} and (c) a non-OCS transmon \mbox{($f_\mathrm{q} = 4.70$ GHz)} with \mbox{$E_\mathrm{J}/E_\mathrm{C} = 74$}. The dashed horizontal lines in (b) and (c) depict the $\overline{T}_\mathrm{1}$ (orange) and $\overline{T}_\mathrm{2, E}$ (blue) of each qubit. Error bars of $T_\mathrm{1}$ and $T_\mathrm{2, E}$ represent the s.e. estimated from the covariance matrix of the fit, while uncertainties of $\overline{T}_\mathrm{1}$ and $\overline{T}_\mathrm{2, E}$ are the s.d. of the time series. Both qubits were measured in the same dilution refrigerator with an identical measurement chain during separate cooldowns (Section S3, Supporting Information). }
\label{figure: qubit}
\end{figure}

The $\mathrm{tan} \ \delta$ of $\beta$-Ta obtained from resonator studies (Figure \ref{figure: tls}b) allows us to compare qubit performance (Figure \ref{figure: summary}g) to the linear absorption limit imposed by surface TLSs. The qubits are designed with a $p_{\mathrm{MS}}$ of $1.3 \times 10^{-4}$, which corresponds to an expected $Q_\mathrm{TLS, 0}$ of 4.8 million based on the extracted $\mathrm{tan} \ \delta$. We note that the distribution of qubit $\overline{Q}$ exceeds the linear absorption limit; this observation is consistent with qubit operation at powers that saturate TLSs \cite{bland2025millisecond, crowley2023disentangling}.

We observe a large spread in $\overline{T}_\mathrm{1}$ and $\overline{T}_\mathrm{2, E}$ across the 11 qubits (Figure \ref{figure: qubit}a). To explain potential reasons underlying this spread, we consider correlations between $T_\mathrm{1}$, $T_\mathrm{2, E}$, and other qubit parameters (Section S4, Supporting Information). We find that the qubits with $\overline{T}_\mathrm{2, E} < \overline{T}_\mathrm{1}$ are also in the offset-charge-sensitive (OCS) regime, as the ratio of their Josephson energy ($E_\mathrm{J}$) to their charging energy ($E_\mathrm{C}$) ranges from $16 - 28$. In contrast, the qubits with $\overline{T}_\mathrm{2, E} > \overline{T}_\mathrm{1}$ are deep in the transmon regime, with an $E_\mathrm{J} \, / \, E_\mathrm{C}$ between $74-122$. By acquiring successive $T_\mathrm{1}$ and $T_\mathrm{2, E}$ measurements over a few days (Figure \ref{figure: qubit}b--c), we calculate a time-averaged pure dephasing time ($\overline{T}_\upphi$), where the pure dephasing time ($T_\upphi$) is obtained using \mbox{$T_\upphi^{-1} = T_\mathrm{2, E}^{-1} - 0.5\, T_\mathrm{1}^{-1}$}. The two groups of qubits exhibit distinct ranges of $\overline{T}_\upphi$, with  $\overline{T}_\upphi$ ranging from $150$ to $470 \ \upmu$s in OCS transmons and $500$ to $1800 \ \upmu$s in non-OCS transmons (Section S4, Supporting Information). The additional dephasing in the OCS transmons likely arises from charge noise, as charge dispersion increases exponentially with decreasing $E_\mathrm{J} \, / \, E_\mathrm{C}$ \cite{koch2007charge}.

We note that qubits with lower $\overline{Q}$ also have higher frequencies (Section S4, Supporting Information), and the qubit with the highest frequency of $5.80$ GHz has the lowest $\overline{Q}$ of $(1.5 \pm 0.2) \times 10^6$, suggesting a frequency-dependent qubit quality factor. We rule out relaxation induced by the Purcell effect \cite{houck2008controlling}, as our devices incorporate an on-chip Purcell filter and electromagnetic simulations indicate a Purcell-limited $T_\mathrm{1}$ exceeding $5$ ms. One possible relaxation channel is radiative coupling to spurious modes of the device package; an electromagnetic simulation of the entire package is needed to verify this possibility \cite{dai2026characterization}.  

Finally, while we use $\overline{T}_\mathrm{1}$ and $\overline{T}_\mathrm{2, E}$ to benchmark qubit performance, the $T_\mathrm{1}$ and $T_\mathrm{2, E}$ distributions for some qubits exhibit strongly non-Gaussian features (Figure \ref{figure: qubit}a). These features indicate temporal fluctuations in $T_\mathrm{1}$ and $T_\mathrm{2, E}$, which likely arise from spectral diffusion of strongly-coupled TLSs within the qubit's frequency neighborhood \cite{klimov2018fluctuations, carroll2022dynamics}. The presence of such non-Gaussian features suggests that the mean and standard deviation of the $T_\mathrm{1}$ and $T_\mathrm{2, E}$ time series may not faithfully represent the variability in qubit performance, and highlights the need for more detailed statistical analysis to benchmark $T_\mathrm{1}$ and $T_\mathrm{2, E}$ fluctuations.

Further improvements to the performance of $\beta$-Ta transmon qubits may be achieved by replacing the sapphire substrate with high-purity silicon, as has been recently demonstrated for $\alpha$-Ta \cite{bland2025millisecond}. We deposit \mbox{$\beta$-Ta} films at room temperature on an intrinsic silicon substrate that is first stripped of its native oxide layer. However, preliminary measurements on CPW resonators fabricated from these films reveal power- and temperature-independent losses that limit $Q_\mathrm{int}$ to $<3$ million across 7 resonators, independent of geometry (Section S10, Supporting Information). Identifying the precise source of this loss requires further careful investigation; one plausible candidate is non-saturable dielectric loss originating from the MS interface \cite{scigliuzzo2020phononic, zhou2025observation}, whose structure differs from that of the $\alpha$-Ta-silicon interface.

This work establishes $\beta$-Ta on sapphire as a material platform for realizing low-loss superconducting quantum devices. The ability to deposit $\beta$-Ta at room temperature makes it a favorable material for industrial-scale Ta qubit fabrication \cite{vandamme2024advanced}. The large kinetic inductance contribution of $\beta$-Ta allows resonators and transmission lines to attain a given frequency at reduced lengths, enabling more compact qubit wiring layouts. Our findings also imply that the presence of minority $\beta$-Ta in $\alpha$-Ta films may not be as detrimental to superconducting circuit performance as previously assumed, and thus, the stringent requirement of depositing ``phase-pure $\alpha$-Ta" can be relaxed. Beyond transmon qubits, the coexistence of high kinetic inductance and high quality factor makes $\beta$-Ta an attractive material for microwave kinetic inductance detectors \cite{kouwenhoven2023resolving, day2003broadband, mazin2022superconducting, chi2024hybrid}. Furthermore, the superconducting gap difference between $\beta$-Ta and other metals such as Al, Nb, or $\alpha$-Ta can be leveraged to realize on-chip QP traps \cite{riwar2019efficient} that protect qubits from relaxation events due to the sudden generation of non-equilibrium QPs \cite{mcewen2024resisting, kurilovich2025correlated}.

\noindent \textbf{Conflict of Interest} \par

Nathalie P. de Leon, Princeton University professor, is also a visiting faculty researcher with Google Quantum AI. As a result of her income from Google, Princeton University has a management plan in place to mitigate a potential conflict of interest that could affect the design, conduct, and reporting of this research. Her academic group also has a sponsored research contract with Google Quantum AI. Andrew A. Houck, Princeton University professor, is also a consultant for Quantum Circuits, Inc. (QCI). As a result of his income from QCI, Princeton University has a management plan in place to mitigate a potential conflict of interest that could affect the design, conduct, and reporting of this research. 

\medskip
\noindent \textbf{Acknowledgments} \par
The authors thank Basil M. Smitham and Saptarshi Chaudhuri for helpful conversations. The authors acknowledge Alexander C. Pakpour-Tabrizi for assisting with qubit fabrication, Ray D. Chang for film growth, and Ambrose Yang and Matthew S. Ai for qubit measurements. This work is primarily supported by the Air Force Office of Scientific Research under award number FA9550-25-1-0172. Some materials characterization is supported by the U.S. Department of Energy, Office of Science, National Quantum Information Science Research Centers, Co-design Center for Quantum Advantage (C2QA) under Contract No. DE-SC0012704. This work was performed in part at the Micro/Nano Fabrication Center (MNFC), a core shared-use facility of the Princeton Materials Institute (PMI). The authors acknowledge the use of Princeton's Imaging and Analysis Center (IAC), which is partially supported by the Princeton Center for Complex Materials (PCCM), a National Science Foundation (NSF) Materials Research Science and Engineering Center (MRSEC; DMR-2011750). The authors acknowledge MIT Lincoln Labs for supplying a traveling-wave parametric amplifier.

\medskip
\noindent \textbf{Data Availability Statement} \par
Data and code associated with this study are available from the corresponding author upon request. 

\raggedright

\bibliographystyle{MSP}
\bibliography{references.bib}

\end{document}


\raggedright

\pagestyle{plain}

\title{Supporting Information for: Beta Tantalum Transmon Qubits with Quality Factors Approaching 10 Million}

\maketitle

\author{Atharv Joshi}
\author{Apoorv Jindal}
\author{Paal H. Prestegaard}
\author{Faranak Bahrami}
\author{Elizabeth Hedrick}
\author{Matthew P. Bland}
\author{Tunmay Gerg}
\author{Guangming Cheng}
\author{Nan Yao}
\author{Robert J. Cava}
\author{Andrew A. Houck}
\author{Nathalie P. de Leon*}

\justifying

\renewcommand{\thesection}{S\arabic{section}}
\renewcommand{\thefigure}{S\arabic{figure}}
\renewcommand{\thetable}{S\arabic{table}}
\renewcommand{\theequation}{S\arabic{equation}}

\section{Film deposition and characterization} \label{sec: film dep}

Beta-phase tantalum ($\beta$-Ta) films are deposited on $1'' \times 1''$ pieces of HEMEX grade c-plane sapphire substrates (Crystal Systems) using a two-chamber DC magnetron sputtering system (AJA Orion 8) equipped with a $99.99\%$ pure tantalum (Ta) target. Before deposition, each substrate is ``piranha-cleaned'' to remove surface contaminants using the following recipe: the substrate is immersed in a 2:1 mixture of $95-98\%$ sulfuric acid and $30\%$ hydrogen peroxide solution for 20 min, then rinsed three times in separate beakers of deionized (DI) water, then rinsed once in isopropanol (IPA), and finally blow dried under nitrogen ($\mathrm{N}_2$) flow. The piranha-cleaned substrate is immediately transferred to the load lock chamber of the sputtering system. Once the pressure of the load lock chamber drops below $10^{-6}$ Torr, the sample is transferred to the deposition chamber. The deposition chamber has a base pressure that ranges between $5 \times 10^{-9}$ to $5 \times 10^{-8}$ Torr. The substrate is heated in situ at $300 \ \degree$C for 1 hour and then allowed to cool to room temperature for around 8 hours. Deposition is carried out in two steps. First, to getter the chamber, Ta is sputtered at room temperature for 15 min with the following parameters: 300 W DC power, 3 mTorr Ar pressure, and 25 sccm Ar flow rate. Next, Ta is sputtered onto the substrate at room temperature with the following parameters: 300 W DC power, 3 mTorr Ar pressure, 25 sccm Ar flow rate, and $1.15''$ throw distance. The deposition parameters result in a film growth rate of around 2 \AA $\ \mathrm{s}^{-1}$, and the deposition time is adjusted according to the desired film thickness. The thickness of each film is determined using atomic force microscopy (AFM) by measuring the vertical height difference between the film surface and the exposed substrate on a device fabricated from that film (Table \ref{table:film_thicknesses}). Films F1 to F14 are used to fabricate resonators (Section \ref{sec: resonator tables}) and films F15 and F16 and used to fabricate qubits (Section \ref{sec: qubit tables}).

\begin{table}[H]
    \centering
    \caption{\textbf{Thicknesses of $\beta$-Ta films used in this work.}}
    \label{table:film_thicknesses}
    \begin{tabular*}{\columnwidth}{c 
    @{\extracolsep{\fill}} c 
    @{\extracolsep{\fill}} c
    @{\extracolsep{\fill}} c
    @{\extracolsep{\fill}} c
    @{\extracolsep{\fill}} c
    @{\extracolsep{\fill}} c
    @{\extracolsep{\fill}} c
    @{\extracolsep{\fill}} c}
        \toprule
        Film & F1 & F2 & F3 & F4 & F5 & F6 & F7 & F8 \\
        Thickness ($\upmu$m) & 0.021 & 0.052 & 0.131 & 0.132 & 0.238 & 0.263 & 0.263 & 0.265 \\
        \midrule
        Film & F9 & F10 & F11 & F12 & F13 & F14 & F15 & F16 \\
        Thickness ($\upmu$m) & 0.475 & 0.480 & 0.480 & 0.480 & 1.09 & 2.00 & 0.260 & 0.262 \\
        \bottomrule
    \end{tabular*}
\end{table}

\begin{figure}[t]
\includegraphics[width=\textwidth]{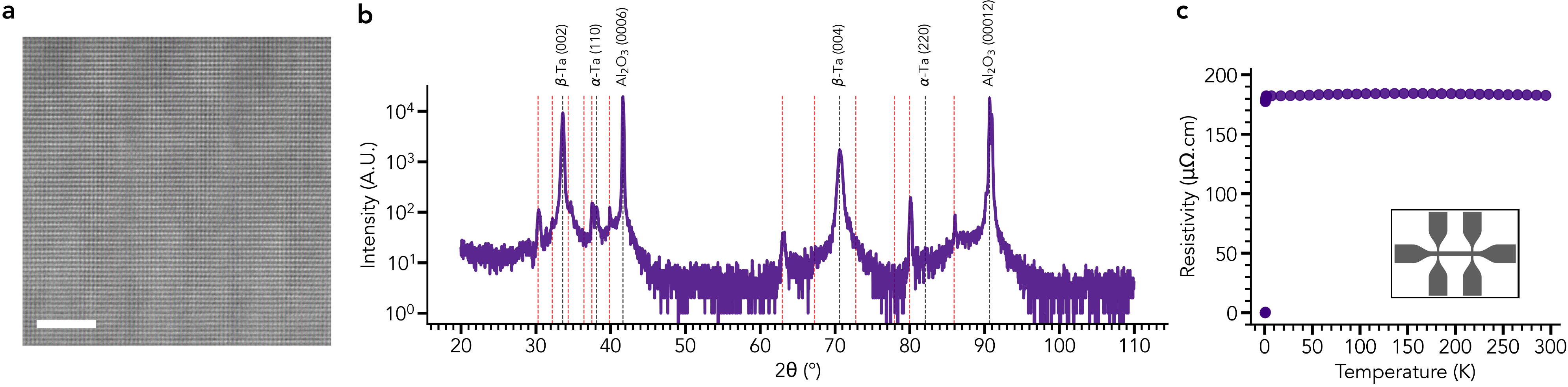}
\caption{\textbf{$\beta$-Ta film characterization.} (a). Cross-sectional STEM image of a Ta film on sapphire depicts a tetragonal crystal structure that is characteristic of $\beta$-Ta. Scale bar: 5 nm. (b) Wide-range XRD spectrum obtained from the $\beta$-Ta film. Dashed black vertical lines indicate the following peak locations: $\beta$-Ta $(002)$ at $33.6 \degree$, $\alpha$-Ta $(110)$ at $38.1 \degree$, $\mathrm{Al_2 O_3} \ (0006)$ at $41.7 \degree$, $\beta$-Ta $(004)$ at $70.6 \degree$, $\alpha$-Ta $(220)$ at $82.1 \degree$, and $\mathrm{Al_2 O_3} \ (00012)$ at $90.7 \degree$. The spectrum reveals $\beta$-Ta as the predominant phase, with traces of $\alpha$-Ta. Dashed red vertical lines indicate artifact peaks arising from Cu $K_\upbeta$ leakage and tungsten contamination. (c) Longitudinal resistivity of a Hall bar device (inset) made from a $\beta$-Ta film measured over the full temperature range of the PPMS.}
\label{figure: si_film_characterization}
\end{figure}

The films are verified to be in the $\beta$ phase using cross-sectional scanning transmission electron microscopy (STEM) (Figure \ref{figure: si_film_characterization}a), X-ray diffraction (XRD) (Figure \ref{figure: si_film_characterization}b), and a four-probe DC resistivity measurement (Figure \ref{figure: si_film_characterization}c). XRD spectra are acquired for all films to ensure that device fabrication (Section \ref{sec: device fab}) proceeds only with predominantly $\beta$-phase films. 

For STEM imaging (FEI Titan Themis$^3$ 300), cross-sectional lamellae are prepared by focused ion beam milling (FEI Helios NanoLab G3 UC DualBeam) and then polished by a 2 keV gallium ion beam to minimize surface damage. Imaging is performed at an accelerating voltage of 300 kV in high-angle annular dark-field (HAADF) STEM mode, with the sample oriented along the $[110]$ zone axis.

XRD patterns are acquired using an X-ray diffractometer (Bruker D8 Advance) equipped with a copper $K_\upalpha$ radiation source and configured with Bragg-Brentano optics. One 0.6 mm slit is inserted before the sample, and one 0.6 mm slit is placed before the detector. The X-ray tube is set to 40 kV and 40 mA. A $2\uptheta$ range of $20\degree-110\degree$ is scanned with a step size of $0.03\degree$ and a speed of $3\degree \ \mathrm{min}^{-1}$. XRD peaks are indexed by comparing the measured diffraction pattern with reference patterns reported in the literature \cite{face1987nucleation, place2021new, crowley2023disentangling}.

For the DC resistivity measurement, a film is patterned into a Hall bar geometry (Figure \ref{figure: si_film_characterization}c inset). Temperature-dependent DC resistance is measured in the longitudinal direction of the Hall bar channel using a four-point configuration in a physical property measurement system (Quantum Design PPMS DynaCool) equipped with a Helium-3 insert. The Hall bar channel length is $370 \ \upmu$m, channel width is $25 \ \upmu$m, and the film thickness is $263$ nm. Measurements are acquired at zero applied magnetic field as the sample is cooled from 295 K down to 0.35 K, with a temperature step size of 10 K in the $5-295$ K range and 0.1 K in $0.35-2.05$ K range. The PPMS stabilizes the temperature at each point before acquiring an average resistance over 25 voltage readings with current-reversal enabled. The normal state resistivity ($\rho_\mathrm{n}$) of the film, reported as the resistivity at 1 K, is $(180 \pm 1) \ \upmu \Omega \ \mathrm{cm}$. The residual resistivity ratio (RRR), calculated as the ratio of the resistivity at 295 K to that at 1 K, is 1.02.

\section{Device fabrication and packaging} \label{sec: device fab}

After deposition, the $\beta$-Ta film is coated with photoresist (AZ 1518) to protect the film surface during the subsequent dicing step. The substrate is diced using a dicing saw (Advanced Dicing Technologies proVectus 7100) into $10 \ \mathrm{mm} \times 10 \ \mathrm{mm}$ chips. The photoresist is removed using the following recipe: the diced chips are immersed in a solvent stripper (Kayaku Remover PG) for 1 hour at $80 \ \degree$C, then sonicated for 2 min each in acetone and IPA, and finally blow dried under $\mathrm{N}_2$ flow. Each chip is piranha-cleaned to remove any residual photoresist using the same recipe described in Section \ref{sec: film dep}. 

The films are patterned into resonators and qubit wiring layers using a subtractive fabrication process. To promote photoresist adhesion during the subsequent photolithography steps, the chip undergoes hexamethyldisilazane vapor priming at $148 \ \degree$C in a vacuum oven (YES Vacuum Oven). AZ 1518 is spun onto the chip at 4000 rpm with a ramp rate of 1000 rpm $\mathrm{s}^{-1}$ for 40 s, and baked at $95 \ \degree$C for 1 min, which results in a $1.9 \ \upmu$m-thick resist layer. The device pattern is exposed onto the chip using a direct-write laser system (Heidelberg DWL66+) with the following configuration: 10 mm write head, 405 nm laser wavelength, 25$\%$ attenuation, 70$\%$ intensity, and optical autofocus mode with $-10 \%$ focus offset. The exposure step is followed by a 2 min bake at 110 $\degree$C. The exposed photoresist is dissolved by immersing the chip in a developer (AZ 300 MIF) for 90 s, followed by rinsing in DI water for 30 s, and finally blow drying under $\mathrm{N_2}$ flow. The device pattern is transferred onto the film using an argon/chlorine-based dry etch in an inductively coupled plasma (ICP) reactive ion etcher (PlasmaTherm Takachi SLR). Before etching the film, the etch chamber is conditioned for 1 hour using the following recipe: 15 sccm Ar flow rate, 20 sccm $\mathrm{Cl}_2$ flow rate, 400 W ICP power, 50 W RF bias power, and 20 mTorr chamber pressure. The etch recipe is as follows: 5 sccm flow rate each of Ar and $\mathrm{Cl_2}$, 500 W ICP power, 50 W RF bias power, and 5.4 mTorr chamber pressure. The resulting etch rate is around 200 nm $\mathrm{min}^{-1}$, and the etch time is adjusted according to the film thickness.

After etching, photoresist is removed using the same recipe described above, followed by another round of piranha cleaning. To reduce the thickness of the native oxide layer, chips are immersed in around 50 mL of 10:1 buffered oxide etchant (Transene) for 20 min, then rinsed three times in separate beakers of DI water, then rinsed once in IPA, and finally blow dried under $\mathrm{N_2}$ flow. At this stage, chips patterned with resonators are ready for packaging, while those with qubit wiring layers proceed for junction fabrication.

For junction fabrication, a bilayer resist stack is spun onto chips patterned with qubit wiring layers. First, a copolymer resist (Kayaku MMA(8.5)MAA EL 13) is spun at 5000 rpm with a ramp rate of 500 rpm s$^{-1}$ for 70 s, and baked at $175 \ \degree$C for 2 min. Next, a poly(methyl methacrylate) resist (Allresist AR-P 672.045 950K) is spun at 4000 rpm with a ramp rate of 500 rpm s$^{-1}$ for 68 s, and baked at $175 \ \degree$C for 5 min. An anti-charging layer (DisChem DisCharge H2O X4) is spun at 4000 rpm with a ramp rate of 1000 rpm s$^{-1}$ for 60 s. Manhattan-style Josephson junction patterns with overlap areas ranging from $0.04-0.18 \ \upmu \mathrm{m^2}$ are exposed onto the chip using an electron beam lithography system (Elionix ELS-F125) operated at 125 kV with a beam current of 1 nA. Patterns are written with a field size of 500 $\upmu$m, a scan/feed pitch of 2.5 nm, and an area dose of 2200 $\upmu \mathrm{C \ cm^{-2}}$. Specific regions of the pattern are exposed with a lower area dose to achieve the required resist undercut for reliable liftoff. After exposure, the anti-charging layer is removed by immersing the chip in DI water for 60 s at room temperature. The exposed pattern is developed by immersing the chip in a 1:3 mixture of DI water and IPA for 150 s, with the developer maintained at $-10 \ \degree$C in a chilled water bath, followed by rinsing in IPA for 15 s at room temperature, and finally blow drying under $\mathrm{N_2}$ flow. The junction overlap areas, exposure settings, and development conditions are determined by fabricating a separate set of test junctions and selecting the parameters that produced junctions with a room-temperature resistance ranging from $5-10 \ \mathrm{k \Omega}$. The double-angle evaporation and oxidation steps used to define the aluminum/aluminum-oxide/aluminum Josephson junctions are performed in an ultrahigh-vacuum electron-beam evaporation system (Plassys MEB550SL3 UHV) following the methods described in Reference \cite{bland2025millisecond} without modification. Finally, the unwanted aluminum is lifted-off by immersing the chip in acetone at room temperature for 12 hours, then sonicating for 1 min each in acetone and IPA, and finally blow drying under $\mathrm{N}_2$ flow.

After device fabrication, the chips are wire-bonded (Questar Q7800) to a printed circuit board that is part of a chip carrier assembly (Quantum Machines QCage.24 with part numbers Q401, Q402-5, Q403\_Al, and Q404\_Al). The chips patterned with qubits include on-chip wire bonds to avoid slotline modes and ensure a continuous microwave ground. The chip carrier assembly is wrapped in a thin layer of aluminized biaxially oriented polyethylene terephthalate (Mylar) to reflect parasitic infrared radiation, mounted on the mixing chamber stage of a dilution refrigerator (Bluefors XLD400 or LD400), and enclosed by a magnetic shield (Quantum Machines QCage Q428). The refrigerator is equipped with an outer magnetic shield that is secured to the mixing chamber stage.

\section{Measurement setup} \label{sec: measurement setup}

\begin{figure}[htbp!]
\fcapside[\FBwidth]
{\caption{\textbf{Dilution refrigerator microwave measurement chain}. Schematic of the microwave components inside the dilution refrigerators used for measuring qubits. The `IN' and `OUT' labels identify the input and output lines respectively, with the arrows indicating the direction of microwave signal flow. The horizontal gray lines labeled `50 K', `4 K', `Still', `Cold plate', and `Mixing chamber' denote the temperature stages of the refrigerator. The triangle labeled `HEMT' represents a high-electron-mobility transistor amplifier (Low Noise Factory LNC4\_8F). Rectangular labels identify the following components: `ECCO', an eccosorb-based infrared filter (Quantum Microwave QMC-CRYOIRF-002); `HERD', a high-energy radiation-drain low-pass filter (Quantum Microwave HERD‑1MF); `HPF', a 6.35 GHz high‑pass filter (Mini-Circuits VHF-6010+); `ISOL', a 22 dB isolator (Low Noise Factory LNF-ISC4\_8A); `LPF', a 12 GHz low‑pass filter (K\&L 6L250-12000/T26000); `QCAGE', the chip carrier assembly described in Section \ref{sec: device fab}; and  `X dB', a cryogenic attenuator with X dB attenuation. The input line contains a total of 66 dB attenuation across all temperature stages. The mixing chamber attenuation chain (top to bottom) is: 10 dB (QMC-CRYOATT), 10 dB (XMA 4880-5523-10-CRYO), 10 dB (QMC-CRYOATT), and 6 dB (QMC-CRYOATT). All attenuators at the other temperature stages are from the XMA 2082-6418-dB-CRYO series. The eccosorb filters and the chip carrier system are enclosed within a magnetic shield (gray cylinder). An outer magnetic shield (not shown) encloses all components mounted on the mixing chamber stage. A traveling wave parametric amplifier (TWPA) (MIT Lincoln Laboratory), enclosed in a dedicated magnetic shield, is used for some qubit measurements. The TWPA is installed at the mixing chamber stage, wired to the output line after the second isolator, and is followed by two more 22 dB LNF-ISC4\_8A isolators. The TWPA pump line is identical to the input line, except for the HERD filter. Resonators are measured using an identical microwave chain, except for the HERD filter.}\label{fig: wiring_chain}}
{\includegraphics[width=0.39\textwidth]{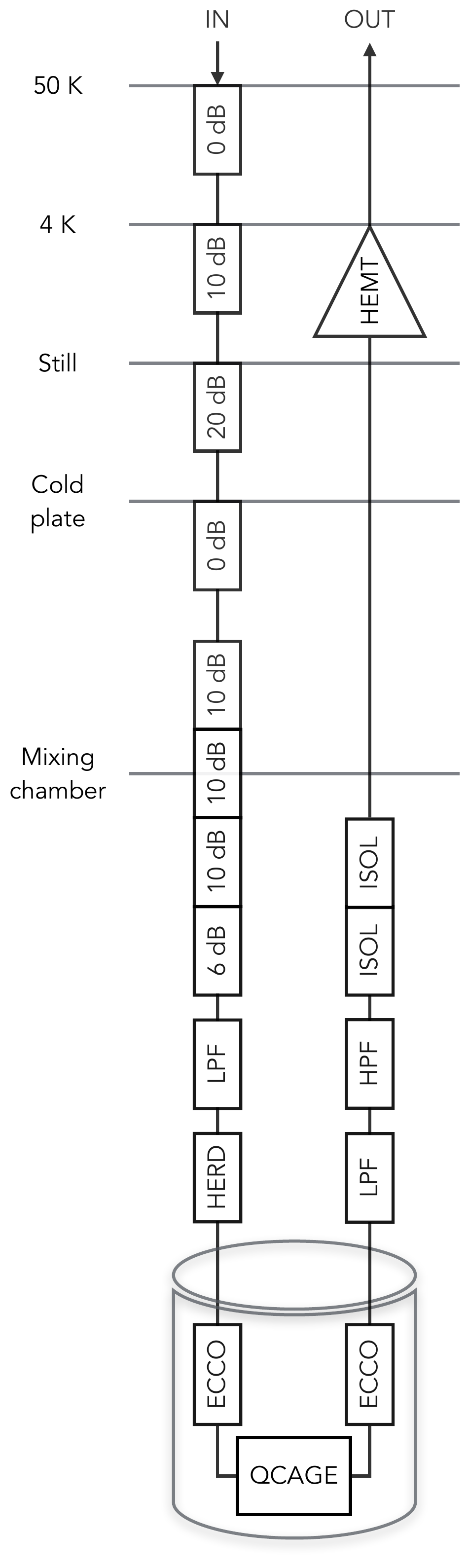}}
\end{figure}

The microwave components installed inside the dilution refrigerators are detailed in Figure \ref{fig: wiring_chain}. 

Resonator transmission measurements (Section \ref{sec: transmission}) are performed with a vector network analyzer (VNA) (Streamline P5008A) at various probe powers and mixing chamber stage temperatures. Probe power is varied using a programmable digital attenuator (Vaunix LDA-608V-4). The total attenuation from the VNA output port to the device input port across $2-8$ GHz is measured separately at room temperature to estimate the power delivered to the device. The output signal from the refrigerator is amplified by a low‑noise amplifier (RF-Lambda RLNA02G08G30) before returning to the VNA. The mixing chamber stage temperature is controlled using the on-stage heater in a closed-loop mode and is stabilized to within 1\% of the target temperature for at least 1 hour before acquiring measurements.

Qubit control and readout pulses are generated at intermediate frequency by separate channels of a quantum control system (Quantum Machines OPX+). The signal from each channel is up-converted using an IQ mixer (Marki MMIQ-0218LSM-2) referenced to its own microwave local oscillator (Agilent E8267D). The up‑converted qubit drive passes through a low‑pass filter (Mini‑Circuits VLF‑5500+), a high‑power amplifier (Mini‑Circuits ZVE‑3W‑83+), and a second VLF‑5500+ before entering the refrigerator input line. The up‑converted readout drive passes through a band‑pass filter (Mini‑Circuits VBF-7331+ or VBF-7900+) before entering the refrigerator input line. The output signal from the refrigerator is amplified by a low‑noise amplifier (RLNA02G08G30), down‑converted with a double-balanced mixer (Mini‑Circuits ZX05‑14‑S+) referenced to the readout channel's local oscillator, and returned to the OPX+, where it is further processed to extract amplitude and phase.

\section{Qubit design parameters, lifetimes, and coherence times} \label{sec: qubit tables}

A total of 11 qubits are measured across 2 chips fabricated from films F15 and F16 (Table \ref{table:film_thicknesses}). Measured design parameters for each qubit are listed in Table \ref{table:qubit_parameters}. The qubit frequency ($\omega_\mathrm{q}/2\pi$) is extracted from Ramsey measurements with an added detuning, the resonator frequency ($\omega_\mathrm{r}/2\pi$) and linewidth ($\kappa/2\pi$) are determined by fitting the measured transmission lineshape to an asymmetrical Lorentzian model, the magnitude of the dispersive shift ($|\chi|/2\pi$) is reported as half of the measured difference in the resonator frequency when the qubit is prepared in the ground versus excited states, and the charging energy ($E_\mathrm{C}$) is estimated as the magnitude of the qubit anharmonicity obtained from spectroscopy of the $e \to f$ transition. The Josephson energy ($E_\mathrm{J}$) is then calculated using the transmon approximation, $\hbar\omega_\mathrm{q} \approx \sqrt{8 E_\mathrm{J} E_\mathrm{C}} - E_\mathrm{C}$. Qubits $1-5$ are in the offset-charge-sensitive (OCS) regime, with $E_\mathrm{J}/E_\mathrm{C}$ ranging from $16-28$, while Qubits $6-11$ are deep in the transmon regime, with $E_\mathrm{J}/E_\mathrm{C}$ ranging from $74-122$.

\begin{table}[H]
    \centering
    \caption{\textbf{Qubit design parameters extracted from measurements.}}
    \label{table:qubit_parameters}
    \begin{tabular*}{\columnwidth}{c 
    @{\extracolsep{\fill}} c 
    @{\extracolsep{\fill}} c
    @{\extracolsep{\fill}} c
    @{\extracolsep{\fill}} c
    @{\extracolsep{\fill}} c
    @{\extracolsep{\fill}} c
    @{\extracolsep{\fill}} c
    @{\extracolsep{\fill}} c}
        \toprule
        Qubit & $\omega_\mathrm{q}/2\pi$ (GHz) & $\omega_\mathrm{r}/2\pi$ (GHz) & $|\chi|/2\pi$ (MHz) & $\kappa/2\pi$ (MHz) & $E_\mathrm{C}/h$ (MHz) & $E_\mathrm{J}/E_\mathrm{C}$ \\
        \midrule
        1 & 2.613 & 7.293 & 0.101 & 0.443 & 252 & 16 \\
        2 & 2.736 & 7.541 & 0.098 & 0.353 & 245 & 19 \\
        3 & 2.799 & 7.366 & 0.034 & 0.499 & 240 & 20 \\
        4 & 2.897 & 6.776 & 0.660 & 0.822 & 238 & 22 \\
        5 & 3.193 & 7.031 & 0.198 & 1.089 & 230 & 28 \\
        6 & 4.696 & 7.750 & 0.258 & 0.488 & 202 & 74 \\
        7 & 4.822 & 7.987 & 0.277 & 0.299 & 202 & 77 \\
        8 & 4.910 & 7.536 & 0.338 & 0.313 & 200 & 82 \\
        9 & 5.145 & 8.233 & 0.319 & 0.914 & 190 & 99 \\
        10 & 5.356 & 8.354 & 0.430 & 0.586 & 198 & 98 \\
        11 & 5.804 & 7.919 & 0.486 & 0.504 & 192 & 122 \\
        \bottomrule
    \end{tabular*}
\end{table}

To benchmark qubit performance, we repeatedly measure the relaxation time ($T_\mathrm{1}$), Ramsey coherence time ($T_\mathrm{2, R}$), and Hahn echo coherence time ($T_\mathrm{2, E}$)  for each qubit over a period of $2-5$ days. For $T_\mathrm{1}$ and $T_\mathrm{2, E}$ measurements, delay times are sampled on a logarithmic scale to ensure robust exponential fits across both short‑ and long‑time regimes. For $T_\mathrm{2, R}$ measurements, an intentional detuning is applied and delay times are linearly spaced; and as the data exhibit beat frequencies, we fit the traces to a sum of cosines under a common exponential envelope to extract the frequencies and $T_\mathrm{2, R}$. Table \ref{table:qubit_performance} reports, for each performance metric, the mean and standard deviation of the corresponding time series in columns $\overline{T}_\mathrm{1}$, $\overline{T}_\mathrm{2, R}$, and $\overline{T}_\mathrm{2, E}$, together with the number of repeated measurements in the columns $N_{T_\mathrm{1}}$, $N_{T_\mathrm{2, R}}$, and $N_{T_\mathrm{2, E}}$  respectively. The table also reports the time-averaged quality factor $\overline{Q} = \omega_\mathrm{q}\overline{T}_\mathrm{1}$, and the time-averaged pure dephasing time $\overline{T}_\upphi$. The $T_\upphi$ time series is constructed from consecutive $(T_\mathrm{1}, \ T_\mathrm{2, E})$ pairs acquired successively using  \mbox{$T_\upphi^{-1} = T_\mathrm{2, E}^{-1} - 0.5\, T_\mathrm{1}^{-1}$}; the number of such pairs is listed as $N_{T_\upphi}$.

\begin{table}[H]
    \centering
    \caption{\textbf{Qubit relaxation and coherence benchmarking results.} $\overline{T}_\upphi$ data is not recorded for Qubit 3.}
    \label{table:qubit_performance}
    \begin{tabular*}{\columnwidth}
    {c@{\extracolsep{\fill}} 
    c@{\extracolsep{\fill}}
    c@{\extracolsep{\fill}}
    c@{\extracolsep{\fill}}
    c@{\extracolsep{\fill}}
    c@{\extracolsep{\fill}}
    c@{\extracolsep{\fill}} 
    c@{\extracolsep{\fill}} 
    c@{\extracolsep{\fill}}c}
        \toprule
        Qubit & $\overline{T}_\mathrm{1}$ ($\upmu$s) & $N_{T_\mathrm{1}}$ & $\overline{Q}$ ($\times 10^6$) & $\overline{T}_\mathrm{2, R}$ ($\upmu$s) & $N_{T_\mathrm{2, R}}$ & $\overline{T}_\mathrm{2, E}$ ($\upmu$s) & $N_{T_\mathrm{2, E}}$ &  $\overline{T}_\upphi$ ($\upmu$s) & $N_{T_\upphi}$ \\
        \midrule
        1 & $397 \, \pm \, 42$ & 2036 & $6.5 \, \pm \, 0.7$ & $53 \, \pm \, 46$ & 295 & $216 \, \pm \, 38$ & 551 & $300 \, \pm \, 71$ & 281 \\
        2 & $585 \, \pm \, 75$ & 731 & $10.1 \, \pm \, 1.3$ & $136 \, \pm \, 67$ & 681 & $328 \, \pm \, 54$ & 753 & $467 \, \pm \, 102$ & 725 \\
        3 & $400 \, \pm \, 68$ & 171 & $7.0 \, \pm \, 1.2$ & $86 \, \pm \, 67$ & 31 & $304 \, \pm \, 52$ & 125 & --- & --- \\
        4 & $423 \, \pm \, 64$ & 680 & $7.7 \, \pm \, 1.2$ & $80 \, \pm \, 57$ & 603 & $123 \, \pm \, 35$ & 742 & $151 \, \pm \, 46$ & 667 \\
        5 & $329 \, \pm \, 29$ & 1449 & $6.6 \, \pm \, 0.6$ & $83 \, \pm \, 21$ & 180 & $159 \, \pm \, 15$ & 533 & $205 \, \pm \, 24$ & 291 \\
        6 & $236 \, \pm \, 42$ & 791 & $7.0 \, \pm \, 1.2$ & $19 \, \pm \, 5$ & 681 & $338 \, \pm \, 69$ & 791 & $1300 \, \pm \, 560$ & 786 \\
        7 & $148 \, \pm \, 11$ & 788 & $4.5 \, \pm \, 0.4$ & $205 \, \pm \, 30$ & 726 & $256 \, \pm \, 23$ & 788 & $1800 \, \pm \, 750$ & 653 \\
        8 & $97 \, \pm \, 13$ & 791 & $3.0 \, \pm \, 0.4$ & $111 \, \pm \, 22$ & 793 & $175 \, \pm \, 25$ & 791 & $1420 \, \pm \, 590$ & 506 \\
        9 & $111 \, \pm \, 11$ & 755 & $3.6 \, \pm \, 0.4$ & $107 \, \pm \, 35$ & 777 & $185 \, \pm \, 26$ & 770 & $1270 \, \pm \, 620$ & 666 \\
        10 & $137 \, \pm \, 11$ & 745 & $4.6 \, \pm \, 0.4$ & $11 \, \pm \, 3$ & 727 & $191 \, \pm \, 45$ & 61 & $506 \, \pm \, 100$ & 6 \\
        11 & $40 \, \pm \, 6$ & 768 & $1.5 \, \pm \, 0.2$ & $34 \, \pm \, 9$ & 788 & $78 \, \pm \, 6$ & 391 & $570 \, \pm \, 166$ & 233 \\
        \bottomrule
    \end{tabular*}
\end{table}

\begin{figure}[H]
\includegraphics[width=\textwidth]{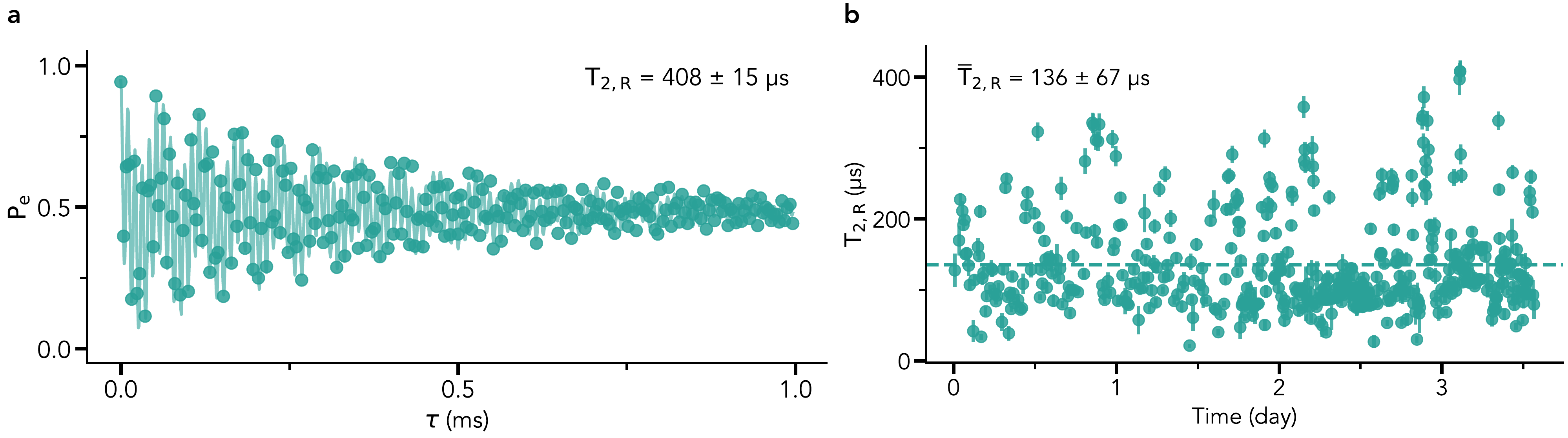}
\caption{\textbf{Ramsey coherence times of Qubit 2.} (a) Excited state population ($P_\mathrm{e}$) as a function of delay time ($\tau$) in a Ramsey experiment showing a maximum $T_\mathrm{2, R} = 408 \pm 15 \ \upmu$s. The fitted beat frequencies are 17 kHz and 95 kHz. The solid line is a fit to the data points. The uncertainty in $T_\mathrm{2, R}$ represents the standard error (s.e.) estimated from the covariance matrix of the fit. (b) Time series of $T_\mathrm{2, R}$ measured continuously over around 3.5 days. The dashed horizontal line indicates the time-averaged value $\overline{T}_\mathrm{2, R}$; the uncertainty in $\overline{T}_\mathrm{2, R}$ is the standard deviation (s.d.) of the time series.}
\label{figure: si_best_t2r}
\end{figure}

\begin{figure}[t]
\includegraphics[width=\textwidth]{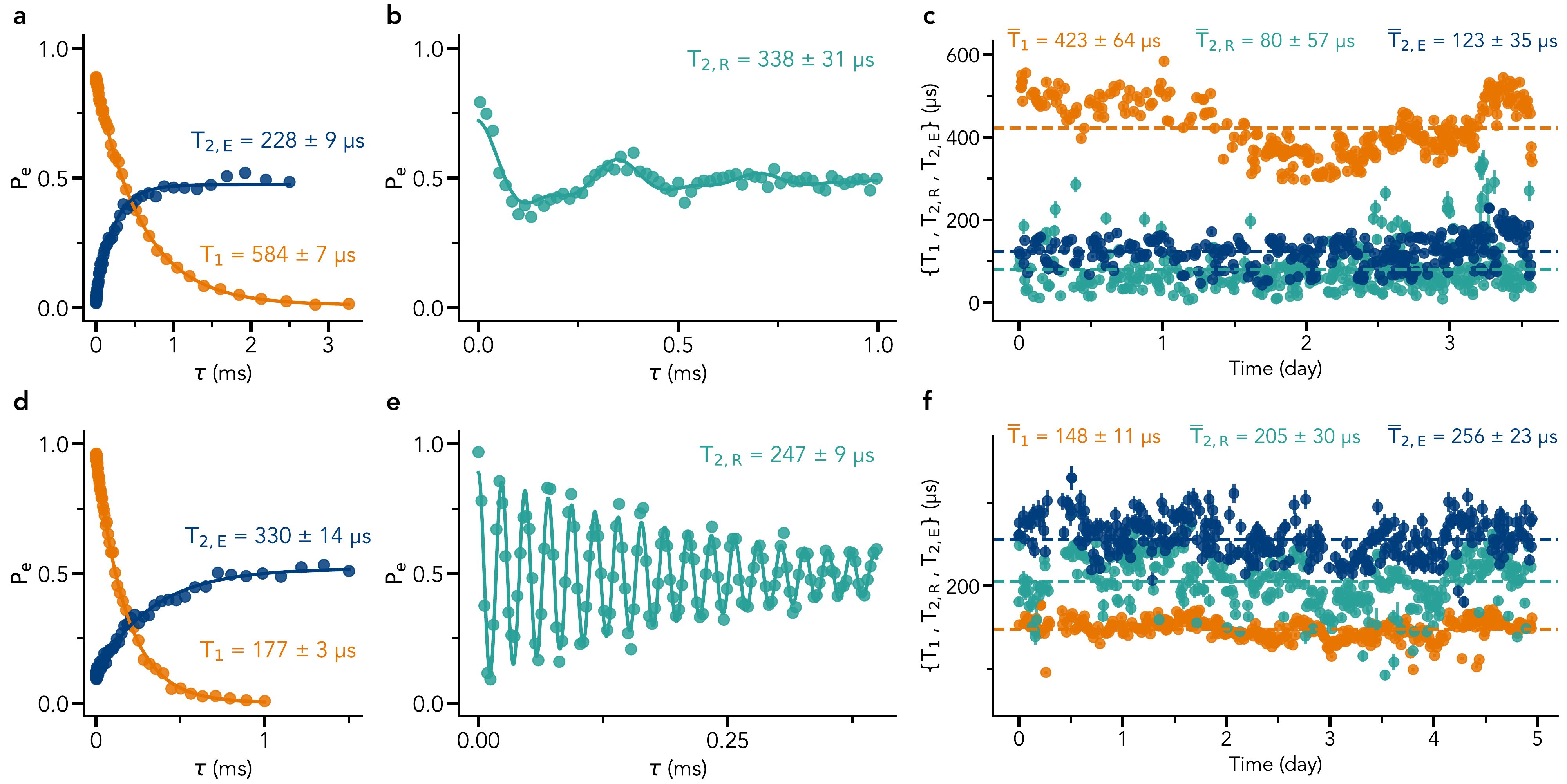}
\caption{\textbf{Relaxation and coherence measurements on Qubit 4 (top row) and Qubit 7 (bottom row).} For each qubit, the panels show: (a, d) Traces corresponding to the maximum measured $T_\mathrm{1}$ (orange) and $T_\mathrm{2, E}$ (blue).  (b, e) Traces corresponding to the maximum measured $T_\mathrm{2, R}$ (teal). In panels a, b, d, and e, solid lines depict fits to the data, and uncertainties in $T_\mathrm{1}$, $T_\mathrm{2, R}$, and $T_\mathrm{2, E}$ represent the s.e. estimated from the covariance matrix of the fit. (c, f) Time series of $T_\mathrm{1}$ (orange), $T_\mathrm{2, R}$ (teal), and $T_\mathrm{2, E}$ (blue) measured continuously over a few days. In panels c and f, the time-averaged values $\overline{T}_\mathrm{1}$, $\overline{T}_\mathrm{2, R}$, and $\overline{T}_\mathrm{2, E}$ are indicated by dashed horizontal lines and their uncertainties are the s.d. of the respective time series.}
\label{figure: si_t1_t2r_t2e}
\end{figure}

Next, we illustrate some additional data to complement the summary statistics in Table \ref{table:qubit_performance}. We highlight the $T_\mathrm{2, R}$ data of the device with the highest $\overline{Q}$, Qubit 2 (Figure \ref{figure: si_best_t2r}). The corresponding $T_\mathrm{1}$ and $T_\mathrm{2, E}$ measurements for Qubit 2 appear in the main text (Figures 1e, 1f, and 4b) and are not repeated here. We also present the $T_\mathrm{1}$, $T_\mathrm{2, R}$, and $T_\mathrm{2, E}$ time-series data for two qubits (Qubits 4 and 7) not included in the main text (Figure \ref{figure: si_t1_t2r_t2e}). Qubit 4 is an OCS transmon ($E_\mathrm{J}/E_\mathrm{C} = 22$) and Qubit 7 is a non-OCS transmon ($E_\mathrm{J}/E_\mathrm{C} = 77$). As discussed in the main text, we measure $\overline{T}_\mathrm{2, E} < \overline{T}_\mathrm{1}$ for OCS transmons and $\overline{T}_\mathrm{2, E} > \overline{T}_\mathrm{1}$ for non-OCS transmons; Qubits 4 and 7 exemplify these respective trends. 

\section{Resonator design parameters, kinetic inductance, and quality factors} \label{sec: resonator tables} 

\begin{figure}[htbp!]
\includegraphics[width=\textwidth]{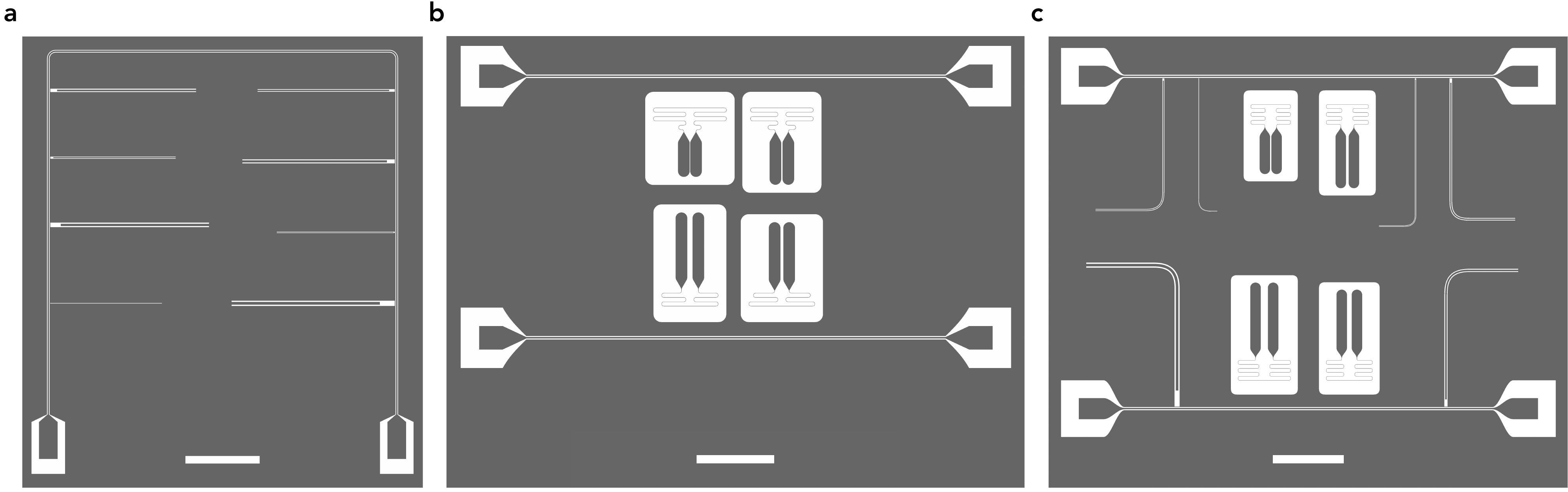}
\caption{\textbf{Resonator chip layouts}. Representative layouts comprising (a) 8 CPW resonators, (b) 4 LE resonators, and (c) 4 LE and 6 CPW resonators. Layouts (a) and (b) were patterned on Film F3 ($t = 131$ nm) and Film F12 ($t = 480$ nm) respectively. Layout (c) was replicated on Films F4 ($t = 132$ nm) , F7 ($t = 263$ nm), and F8 ($t = 265$ nm). All scale bars represent 1 mm.}
\label{figure: resonator_designs}
\end{figure}

A total of 80 coplanar waveguide (CPW) resonators and 16 lumped element (LE) resonators are measured on 14 chips fabricated from films with different thicknesses ($\mathrm{F1}-\mathrm{F14}$; Table \ref{table:film_thicknesses}). Initial measurements showed significantly lower resonance frequencies compared to those obtained from electromagnetic simulations that compute only the capacitive and inductive contributions from the resonator geometry, suggesting an additional kinetic inductance contribution (Figure 2a, Main Text). Using these initial results, we estimated the expected sheet kinetic inductance ($L_\mathrm{k/ \square}$) for each subsequent film thickness ($t$) and incorporated it into the simulations to predict resonance frequencies including both geometric and kinetic inductance contributions (Section \ref{sec: simulations}). We then scaled the CPW center conductor length or the LE meander-inductor length of each resonator to place the fundamental resonance frequency within the $3.5-8.5$ GHz operating band. As a result, films with smaller thicknesses were patterned with resonators that had shorter inductive elements; three representative resonator chip layouts are illustrated in Figure \ref{figure: resonator_designs}. Moreover, for a given film thickness, as the kinetic inductance fraction of CPW resonators increases with decreasing gap width, we arranged their target frequencies in ascending order of gap width; this monotonic ordering ensures unambiguous resonator identification during microwave loss measurements. 

\afterpage{
\clearpage
\begin{longtable}
    {l@{\extracolsep{\fill}} 
    c@{\extracolsep{\fill}}
    c@{\extracolsep{\fill}}
    c@{\extracolsep{\fill}}
    c@{\extracolsep{\fill}}
    c@{\extracolsep{\fill}}
    c@{\extracolsep{\fill}} 
    c@{\extracolsep{\fill}} 
    c@{\extracolsep{\fill}}c}
    \caption{\textbf{Summary of resonator measurements.} $L_\mathrm{k/\square}$ is not computed for LE resonators made from Films F10 and F13. $Q_\mathrm{TLS, 0}$ and $Q_\mathrm{other}$ are omitted where power-temperature sweeps were not performed or fitted values had large errors.}    
    \label{table:resonators}
    \endfirsthead
    \toprule
    Resonator & $p_\mathrm{MS} \ (\times10^{-4})$ & $|Q_\mathrm{c}| \ (\times10^6)$ & $f_\mathrm{r, m}$ (GHz) & $\alpha$ & $L_\mathrm{k/\square} \ (\mathrm{pH/\square})$ & $Q_\mathrm{TLS, 0} \ (\times10^6)$ & $Q_\mathrm{other} \ (\times10^6)$ \\
    \midrule
    \endhead
    \toprule
    Resonator & $p_\mathrm{MS} \ (\times10^{-4})$ & $|Q_\mathrm{c}| \ (\times10^6)$ & $f_\mathrm{r, m}$ (GHz) & $\alpha$ & $L_\mathrm{k/\square} \ (\mathrm{pH/\square})$ & $Q_\mathrm{TLS, 0} \ (\times10^6)$ & $Q_\mathrm{other} \ (\times10^6)$ \\
    \midrule
        F1--CPW--6 & 8.63 & 0.72 & 4.151 & 0.978 & 244 $\pm$ 31 & --- & --- \\
        F1--CPW--8 & 6.72 & 0.13 & 4.465 & 0.971 & 259 $\pm$ 33 & --- & --- \\
        F1--CPW--10 & 5.53 & 0.04 & 4.777 & 0.964 & 239 $\pm$ 30 & --- & --- \\
        F1--CPW--12 & 4.72 & 0.22 & 4.953 & 0.958 & 261 $\pm$ 33 & --- & --- \\
        F1--CPW--14 & 4.13 & 0.13 & 5.183 & 0.950 & 226 $\pm$ 29 & --- & --- \\
        F1--CPW--16 & 3.68 & 0.27 & 5.348 & 0.945 & 246 $\pm$ 32 & --- & --- \\
        F2--CPW--2 & 22.07 & 2.75 & 4.241 & 0.958 & 40 $\pm$ 5 & 0.72 $\pm$ 0.14 & 5.58 $\pm$ 2.36 \\
        F2--CPW--4 & 12.24 & 1.85 & 4.537 & 0.947 & 64 $\pm$ 8 & 0.30 $\pm$ 0.06 & 1.93 $\pm$ 0.30 \\
        F2--CPW--6 & 8.63 & 8.80 & 4.774 & 0.935 & 79 $\pm$ 10 & 0.85 $\pm$ 0.28 & 3.00 $\pm$ 0.34 \\
        F2--CPW--8 & 6.72 & 0.18 & 5.126 & 0.913 & 74 $\pm$ 10 & --- & --- \\
        F2--CPW--10 & 5.53 & 0.07 & 5.288 & 0.896 & 74 $\pm$ 10 & --- & --- \\
        F2--CPW--12 & 4.72 & 17.7 & 5.547 & 0.875 & 71 $\pm$ 10 & 1.02 $\pm$ 0.18 & 4.22 $\pm$ 0.62 \\
        F2--CPW--14 & 4.13 & 1.73 & 5.718 & 0.861 & 74 $\pm$ 10 & 1.15 $\pm$ 0.31 & 3.68 $\pm$ 0.70 \\
        F2--CPW--16 & 3.67 & 4.76 & 6.077 & 0.840 & 72 $\pm$ 10 & 1.05 $\pm$ 0.20 & 2.85 $\pm$ 0.19 \\
        F3--CPW--2 & 22.07 & 1.86 & 4.391 & 0.946 & 34 $\pm$ 4 & 0.38 $\pm$ 0.06 & $>10^5$ \\
        F3--CPW--4 & 12.24 & 0.95 & 5.714 & 0.900 & 33 $\pm$ 5 & 0.56 $\pm$ 0.16 & 7.50 $\pm$ 2.46 \\
        F3--CPW--8 & 6.72 & 0.75 & 6.826 & 0.821 & 32 $\pm$ 5 & 0.41 $\pm$ 0.05 & 3.32 $\pm$ 0.44 \\
        F3--CPW--10 & 5.53 & 0.49 & 7.005 & 0.791 & 32 $\pm$ 5 & --- & --- \\
        F3--CPW--12 & 4.72 & 7.76 & 7.206 & 0.760 & 38 $\pm$ 6 & 0.88 $\pm$ 0.31 & 4.30 $\pm$ 1.04 \\
        F3--CPW--16 & 3.67 & 3.80 & 7.716 & 0.711 & 31 $\pm$ 5 & 2.69 $\pm$ 0.48 & 3.27 $\pm$ 0.11 \\
        F4--CPW--4 & 12.24 & 0.50 & 3.880 & 0.903 & 34 $\pm$ 5 & 0.64 $\pm$ 0.14 & $>15$ \\
        F4--CPW--6 & 8.63 & 1.23 & 4.290 & 0.864 & 33 $\pm$ 5 & 0.77 $\pm$ 0.22 & $>11$ \\
        F4--CPW--8 & 6.72 & 0.57 & 4.742 & 0.828 & 33 $\pm$ 5 & 1.29 $\pm$ 0.28 & 3.03 $\pm$ 0.31 \\
        F4--CPW--10 & 5.53 & 1.37 & 5.210 & 0.796 & 32 $\pm$ 5 & 1.53 $\pm$ 0.29 & 2.37 $\pm$ 0.14 \\
        F4--CPW--16 & 3.67 & 0.11 & 5.656 & 0.717 & 33 $\pm$ 5 & --- & --- \\
        F5--CPW--4 & 12.24 & 0.79 & 3.146 & 0.818 & 15 $\pm$ 2 & --- & --- \\
        F5--CPW--6 & 8.63 & 0.22 & 3.510 & 0.792 & 19 $\pm$ 3 & 1.33 $\pm$ 0.24 & 2.29 $\pm$ 0.14 \\
        F5--CPW--8 & 6.72 & 5.75 & 3.913 & 0.758 & 20 $\pm$ 3 & 0.76 $\pm$ 0.06 & $>10^6$ \\
        F5--CPW--10 & 5.53 & 4.14 & 3.386 & 0.673 & 16 $\pm$ 3 & 0.63 $\pm$ 0.19 & 3.77 $\pm$ 1.65 \\
        F5--CPW--14 & 4.13 & 27.6 & 3.833 & 0.652 & 19 $\pm$ 3 & 0.81 $\pm$ 0.21 & 12.0 $\pm$ 5.0 \\
        F5--CPW--16 & 3.67 & 33.5 & 4.161 & 0.623 & 19 $\pm$ 3 & 0.52 $\pm$ 0.24 & $>10^5$ \\
        F6--CPW--2 & 22.07 & 3.22 & 6.012 & 0.845 & 10 $\pm$ 1 & 0.95 $\pm$ 0.16 & 2.61 $\pm$ 0.23 \\
        F6--CPW--6 & 8.63 & 1.00 & 6.812 & 0.658 & 10 $\pm$ 2 & 0.46 $\pm$ 0.12 & 3.07 $\pm$ 2.30 \\
        F6--CPW--8 & 6.72 & 0.77 & 7.147 & 0.608 & 11 $\pm$ 2 & --- & --- \\
        F6--CPW--10 & 5.53 & 0.35 & 7.406 & 0.587 & 12 $\pm$ 2 & --- & --- \\
        F6--CPW--16 & 3.67 & 0.69 & 7.708 & 0.474 & 12 $\pm$ 3 & --- & --- \\
        F6--LE--15 & 2.36 & 3.42 & 5.185 & 0.752 & 11 $\pm$ 2 & 7.34 $\pm$ 1.41 & 16.8 $\pm$ 2.6 \\
        F6--LE--35 & 1.42 & 5.69 & 4.680 & 0.758 & 13 $\pm$ 2 & 5.89 $\pm$ 0.60 & 21.0 $\pm$ 2.5 \\
        F6--LE--65 & 1.01 & 5.45 & 3.997 & 0.808 & 16 $\pm$ 2 & 9.92 $\pm$ 6.94 & 18.9 $\pm$ 8.1 \\
        F6--LE--100 & 0.81 & 4.54 & 3.779 & 0.803 & 16 $\pm$ 2 & 7.41 $\pm$ 0.78 & 17.0 $\pm$ 1.4 \\
        F7--CPW--2 & 22.07 & 5.16 & 4.594 & 0.907 & 18 $\pm$ 2 & 0.27 $\pm$ 0.04 & 9.02 $\pm$ 3.59 \\
        F7--CPW--4 & 12.24 & 0.75 & 4.998 & 0.835 & 18 $\pm$ 3 & 0.53 $\pm$ 0.07 & $>30$ \\
        F7--CPW--6 & 8.63 & 0.43 & 5.591 & 0.769 & 17 $\pm$ 3 & 1.08 $\pm$ 0.10 & 15.2 $\pm$ 2.6 \\
        F7--CPW--8 & 6.72 & 0.60 & 5.969 & 0.725 & 17 $\pm$ 3 & 0.99 $\pm$ 0.16 & 4.79 $\pm$ 0.56 \\
        F7--CPW--10 & 5.53 & 0.24 & 6.553 & 0.675 & 17 $\pm$ 3 & 1.34 $\pm$ 0.22 & 1.53 $\pm$ 0.02 \\
        F7--CPW--12 & 4.72 & 0.26 & 6.917 & 0.643 & 17 $\pm$ 3 & 1.00 $\pm$ 0.66 & 8.51 $\pm$ 0.85 \\
        F7--CPW--14 & 4.13 & 0.07 & 6.921 & 0.628 & 19 $\pm$ 3 & 1.53 $\pm$ 0.86 & 4.91 $\pm$ 0.54 \\
        F7--CPW--16 & 3.67 & 0.10 & 7.339 & 0.586 & 17 $\pm$ 3 & 3.89 $\pm$ 2.09 & 3.86 $\pm$ 0.18 \\
        F8--CPW--2 & 22.07 & 2.38 & 4.552 & 0.911 & 19 $\pm$ 3 & 0.28 $\pm$ 0.05 & $>10^5$ \\
        F8--CPW--4 & 12.24 & 1.41 & 5.026 & 0.838 & 19 $\pm$ 3 & 0.77 $\pm$ 0.15 & 12.8 $\pm$ 6.7 \\
        F8--CPW--6 & 8.63 & 0.73 & 5.476 & 0.779 & 18 $\pm$ 3 & 0.88 $\pm$ 0.16 & 17.4 $\pm$ 11.2 \\
        F8--CPW--8 & 6.72 & 1.10 & 5.953 & 0.728 & 18 $\pm$ 3 & 1.35 $\pm$ 0.37 & 6.20 $\pm$ 1.41 \\
        F8--CPW--16 & 3.67 & 0.39 & 6.832 & 0.587 & 19 $\pm$ 4 & 0.40 $\pm$ 0.19 & 1.18 $\pm$ 0.07 \\
        F8--LE--15 & 2.36 & 5.79 & 4.164 & 0.840 & 20 $\pm$ 3 & 2.96 $\pm$ 0.66 & 12.7 $\pm$ 2.0 \\
        F8--LE--35 & 1.42 & 5.91 & 3.847 & 0.837 & 20 $\pm$ 3 & 4.58 $\pm$ 2.35 & 14.7 $\pm$ 4.9 \\
        F8--LE--65 & 1.01 & 4.84 & 3.785 & 0.828 & 19 $\pm$ 3 & 3.08 $\pm$ 1.79 & $>30$ \\
        F8--LE--100 & 0.81 & 3.51 & 3.532 & 0.828 & 19 $\pm$ 3 & --- & --- \\
        F9--CPW--4 & 12.24 & 2.34 & 3.585 & 0.764 & 11 $\pm$ 2 & 0.66 $\pm$ 0.23 & $>200$ \\
        F9--CPW--6 & 8.63 & 1.79 & 4.313 & 0.686 & 11 $\pm$ 2 & 0.82 $\pm$ 0.13 & $>300$ \\
        F9--CPW--8 & 6.72 & 7.58 & 4.889 & 0.622 & 10 $\pm$ 2 & 1.17 $\pm$ 0.54 & $>10^3$ \\
        F9--CPW--10 & 5.53 & 0.31 & 3.870 & 0.573 & 10 $\pm$ 2 & 1.07 $\pm$ 0.46 & 3.20 $\pm$ 0.17 \\
        F9--CPW--12 & 4.72 & 4.09 & 4.260 & 0.531 & 10 $\pm$ 2 & 1.48 $\pm$ 0.28 & $>90$ \\
        F9--CPW--14 & 4.13 & 7.20 & 4.601 & 0.498 & 10 $\pm$ 2 & 1.34 $\pm$ 0.45 & 31 $\pm$ 12 \\
        F9--CPW--16 & 3.67 & 5.78 & 4.949 & 0.467 & $10 \pm 2$ & 0.98 $\pm$ 0.30 & 26 $\pm$ 5 \\
        F10--LE--5 & 4.84 & 2.73 & 4.477 & 0.753 & --- & 3.43 $\pm$ 0.57 & 3.20 $\pm$ 0.16 \\
        F10--LE--15 & 2.36 & 3.29 & 4.669 & 0.755 & --- & 6.30 $\pm$ 1.42 & 3.63 $\pm$ 0.13 \\
        F10--LE--35 & 1.42 & 1.89 & 5.012 & 0.744 & --- & 4.76 $\pm$ 0.48 & 3.98 $\pm$ 0.17 \\
        F10--LE--65 & 1.01 & 1.29 & 5.286 & 0.733 & --- & 5.39 $\pm$ 0.39 & 4.91 $\pm$ 0.26 \\
        F11--LE--5 & 4.84 & 14.21 & 4.057 & 0.797 & --- & 4.89 $\pm$ 1.28 & $>30$ \\
        F11--LE--15 & 2.36 & 5.74 & 4.329 & 0.789 & --- & 4.67 $\pm$ 1.50 & $>75$ \\
        F11--LE--35 & 1.42 & 7.36 & 4.642 & 0.781 & --- & 3.23 $\pm$ 0.50 & 19.3 $\pm$ 3.1 \\
        F11--LE--65 & 1.01 & 4.79 & 4.982 & 0.763 & --- & 4.08 $\pm$ 2.06 & $>10^6$ \\
        F12--CPW--2 & 22.07 & 1.42 & 7.126 & 0.847 & 9.8 $\pm$ 1.4 & 0.45 $\pm$ 0.09 & 4.99 $\pm$ 3.29 \\
        F12--CPW--4 & 12.24 & 2.27 & 3.792 & 0.735 & 9.3 $\pm$ 1.5 & 0.56 $\pm$ 0.12 & $>30$ \\
        F12--CPW--6 & 8.63 & 1.61 & 6.829 & 0.658 & 9.5 $\pm$ 1.6 & 0.71 $\pm$ 0.09 & 5.02 $\pm$ 1.76 \\
        F12--CPW--8 & 6.72 & 2.84 & 6.386 & 0.591 & 9.0 $\pm$ 1.7 & 0.34 $\pm$ 0.21 & 1.77 $\pm$ 0.18 \\
        F12--CPW--10 & 5.53 & 3.20 & 5.993 & 0.543 & 9.2 $\pm$ 1.9 & 1.35 $\pm$ 0.18 & 2.51 $\pm$ 0.11 \\
        F12--CPW--12 & 4.72 & 2.93 & 5.670 & 0.503 & 9.0 $\pm$ 1.9 & 2.24 $\pm$ 0.27 & 8.04 $\pm$ 0.79 \\
        F12--CPW--14 & 4.13 & 2.45 & 5.355 & 0.465 & 8.7 $\pm$ 2.0 & 1.82 $\pm$ 0.28 & 24.1 $\pm$ 13.9 \\
        F12--CPW--16 & 3.67 & 4.64 & 5.087 & 0.435 & 8.9 $\pm$ 2.2 & 1.70 $\pm$ 0.11 & 25.3 $\pm$ 3.2 \\
        F13--CPW--2 & 22.07 & 4.88 & 4.913 & 0.720 & 4.6 $\pm$ 0.7 & --- & --- \\
        F13--CPW--4 & 12.24 & 3.09 & 5.669 & 0.555 & 3.9 $\pm$ 0.8 & 0.48 $\pm$ 0.08 & 1.92 $\pm$ 0.11 \\
        F13--CPW--6 & 8.63 & 2.83 & 6.045 & 0.476 & 4.0 $\pm$ 0.9 & 0.54 $\pm$ 0.25 & 2.45 $\pm$ 0.44 \\
        F13--CPW--8 & 6.72 & 4.28 & 6.401 & 0.403 & 3.8 $\pm$ 1.0 & --- & --- \\
        F13--CPW--10 & 5.53 & 2.66 & 6.614 & 0.362 & 3.9 $\pm$ 1.1 & 0.80 $\pm$ 0.22 & 3.61 $\pm$ 0.45 \\
        F13--CPW--12 & 4.72 & 2.45 & 6.936 & 0.317 & 3.6 $\pm$ 1.2 & 1.76 $\pm$ 0.30 & 5.52 $\pm$ 0.22 \\
        F13--CPW--14 & 4.13 & 1.17 & 7.181 & 0.291 & 3.7 $\pm$ 1.3 & 1.29 $\pm$ 0.24 & 6.32 $\pm$ 1.40 \\
        F13--CPW--16 & 3.67 & 2.86 & 7.550 & 0.255 & 3.5 $\pm$ 1.4 & 1.91 $\pm$ 0.66 & 12.3 $\pm$ 2.2 \\
        F14--CPW--2 & 22.07 & 7.95 & 4.451 & 0.624 & 3.0 $\pm$ 0.5 & --- & --- \\
        F14--CPW--4 & 12.24 & 2.05 & 5.477 & 0.448 & 2.7 $\pm$ 0.6 & --- & --- \\
        F14--CPW--6 & 8.63 & 4.31 & 6.165 & 0.359 & 2.7 $\pm$ 0.8 & --- & --- \\
        F14--CPW--8 & 6.72 & 4.03 & 6.671 & 0.296 & 2.6 $\pm$ 0.9 & --- & --- \\
        F14--CPW--10 & 5.53 & 2.01 & 5.113 & 0.255 & 2.6 $\pm$ 1.0 & --- & --- \\
        F14--CPW--12 & 4.72 & 4.07 & 5.474 & 0.225 & 2.6 $\pm$ 1.2 & --- & --- \\
        F14--CPW--14 & 4.13 & 5.02 & 5.799 & 0.203 & 2.6 $\pm$ 1.3 & --- & --- \\
        F14--CPW--16 & 3.67 & 2.78 & 6.132 & 0.181 & 2.5 $\pm$ 1.4 & --- & --- \\
        \bottomrule
\end{longtable}
\clearpage
}

Table \ref{table:resonators} summarizes key parameters of all 96 resonators. Devices are labeled $\mathrm{F \langle n \rangle}-\mathrm{type}-s$, where $\mathrm{F \langle n \rangle}$ (F1--F14) identifies the metal film the resonator is patterned on (Table \ref{table:film_thicknesses}), type $\in$ \{CPW, LE\} denotes the resonator type, and $s$ is the nominal gap width for CPW resonators and the nominal distance between capacitor pads for LE resonators in $\upmu$m. The surface participation ratio of the metal-substrate interface, $p_\mathrm{MS}$, is computed from electromagnetic simulations (Section \ref{sec: surface_tls_loss}). The coupling quality factor,  $|Q_\mathrm{c}|$, and the measured resonance frequency, $f_\mathrm{r, m}$, are extracted from transmission measurements (Section \ref{sec: transmission}). The effective kinetic inductance fraction $\alpha$ is calculated using $\alpha = 1 - (f_\mathrm{r, m} / f_\mathrm{r, s})^2$, where $f_\mathrm{r, s}$ is the resonance frequency obtained from geometric electromagnetic simulations that neglect kinetic inductance. The sheet kinetic inductance, $L_\mathrm{k/\square}$, is estimated using electromagnetic simulations of the resonator surface current distributions in the presence of finite surface impedance (Section \ref{sec: simulations}). $Q_\mathrm{TLS, 0}$, the inverse linear absorption due to two-level systems (TLS), and $Q_\mathrm{other}$, the quality factor associated with power- and temperature-independent losses, are determined from power and temperature sweeps of the resonator internal quality factor ($Q_\mathrm{int}$) (Section \ref{sec: microwave_losses}). At millikelvin temperatures, $Q_\mathrm{TLS, 0}$ parametrizes the TLS loss contribution that limits $Q_\mathrm{int}$ in the low-power regime, and $Q_\mathrm{other}$ captures the high-power asymptote of $Q_\mathrm{int}$. 

\begin{figure}[htbp!]
\includegraphics[width=\textwidth]{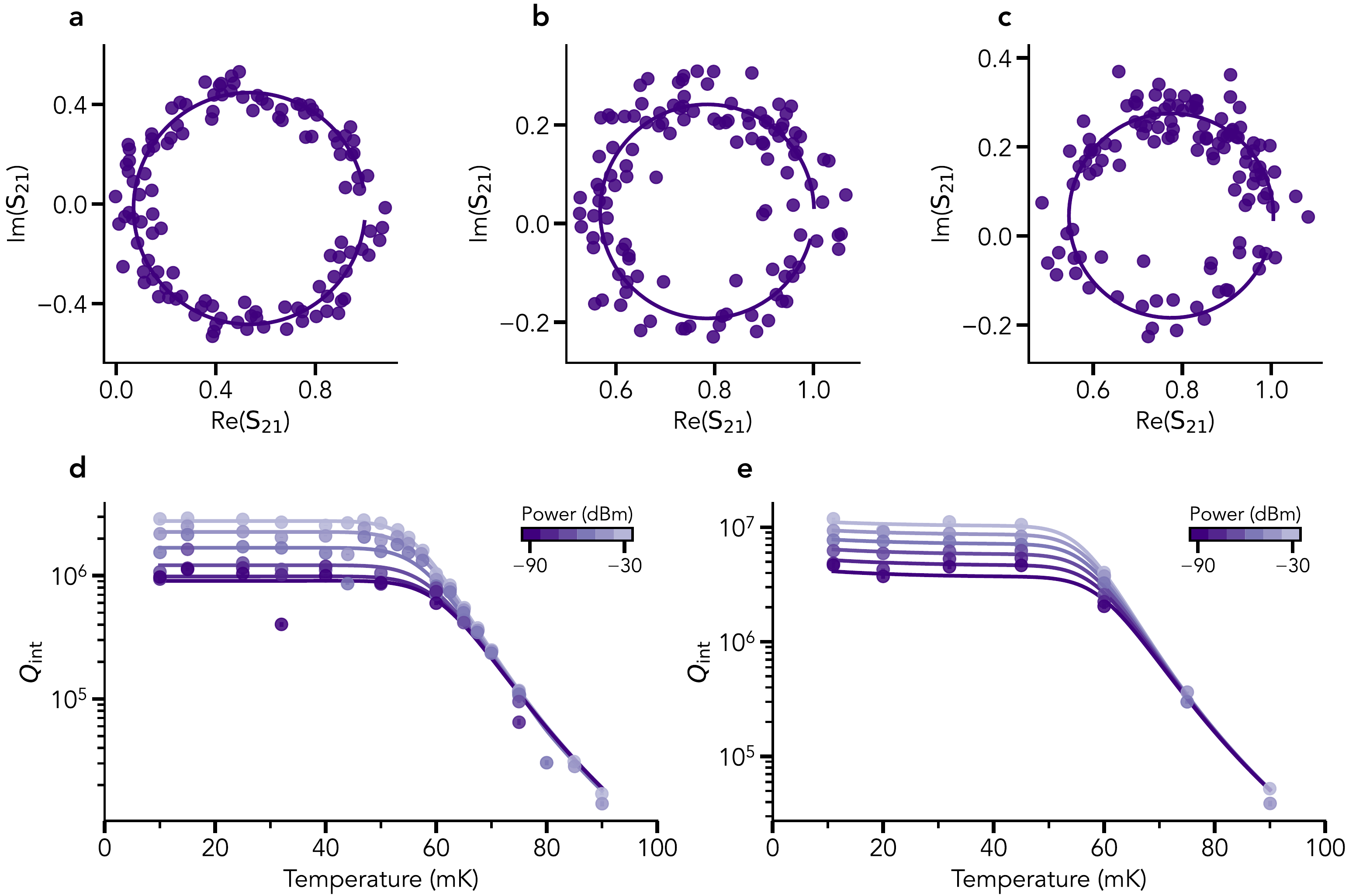}
\caption{\textbf{Coexistence of high \mbox{$L_\mathrm{k/\square}$} and \mbox{$Q_\mathrm{int}$} in $\beta$-Ta resonators.} Real and imaginary components of the transmission coefficient ($S_{\mathrm{21}}$) with circle fits (solid lines) for (a) a CPW resonator (F1--CPW--8) with \mbox{$L_\mathrm{k/\square} = (259 \pm 33) \ \mathrm{pH/\square}$} and \mbox{$Q_\mathrm{int} = (1.6 \pm 0.1) \times 10^6$} at \mbox{$P_\mathrm{in} = -90$ dBm} and \mbox{$T = 10$ mK}; (b) a CPW resonator (F2--CPW--14) with \mbox{$L_\mathrm{k/\square} = (74 \pm 10) \ \mathrm{pH/\square}$} and \mbox{$Q_\mathrm{int} = (1.1 \pm 0.3) \times 10^6$} at \mbox{$P_\mathrm{in} = -90$ dBm} and \mbox{$T = 15$ mK}; and (c) an LE resonator (F8--LE--35) with \mbox{$L_\mathrm{k/\square} = (20 \pm 3) \ \mathrm{pH/\square}$} and \mbox{$Q_\mathrm{int} = (4.8 \pm 0.1) \times 10^6$} at \mbox{$P_\mathrm{in} = -80$ dBm} and \mbox{$T = 11$ mK}. Panels (d) and (e) show the power- and temperature-dependence of $Q_\mathrm{int}$ with fits (solid lines) to the loss model described in Section \ref{sec: microwave_losses} for the devices in (b) and (c) respectively.}
\label{figure: resonator_qi_lk}
\end{figure}

\begin{figure}[htbp!]
\includegraphics[width=\textwidth]{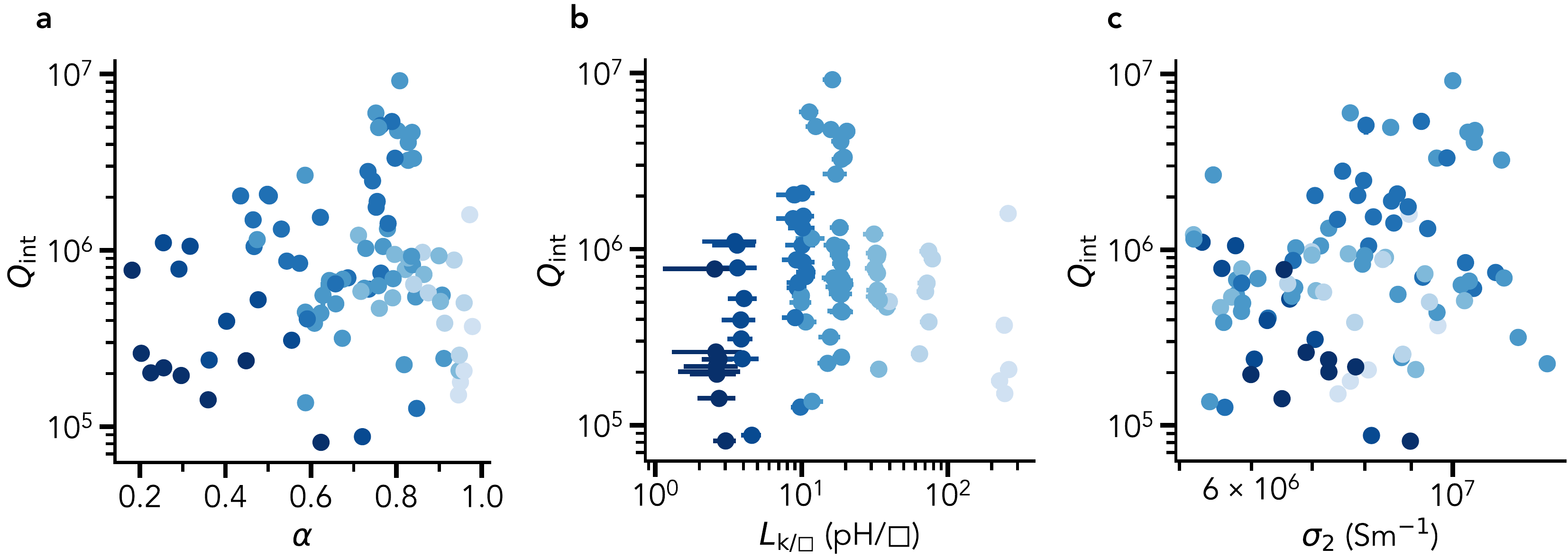}
\caption{\textbf{Relating $Q_\mathrm{int}$ and kinetic inductance metrics across $\beta$-Ta resonators.} The highest $Q_\mathrm{int}$ measured at low nominal input powers ($P_\mathrm{in} < -80$ dBm) and temperatures ($T < 15$ mK) for all devices, plotted versus the (a) effective kinetic inductance fraction $\alpha$, (b) sheet kinetic inductance $L_\mathrm{k/\square}$, and (c) inductive response $\sigma_2$. Shades of blue encode film thickness ($t$) (lightest blue: $t = 0.02 \ \upmu$m; darkest blue: $t = 2.00 \ \upmu$m).}
\label{figure: inductive_losses}
\end{figure}

As noted in the main text, $\beta$-Ta resonators exhibit large $L_\mathrm{k/\square}$ together with high $Q_\mathrm{int}$, a combination advantageous for fabricating highly sensitive microwave kinetic inductance detectors \cite{mazin2022superconducting}. We present transmission measurements and power-temperature sweeps of $Q_\mathrm{int}$ that exemplify high $L_\mathrm{k/\square}$ alongside $Q_\mathrm{int} > 10^6$ (Figure \ref{figure: resonator_qi_lk}). To illustrate the coexistence of high $L_\mathrm{k/\square}$ and $Q_\mathrm{int}$ across the full dataset, we plot the $Q_\mathrm{int}$ measured at low nominal input powers ($P_\mathrm{in}$) and millikelvin temperatures ($T$) versus $\alpha$ (Figure \ref{figure: inductive_losses}a) and $L_\mathrm{k/\square}$ (Figure \ref{figure: inductive_losses}b). 

Finally, we relate $Q_\mathrm{int}$ to the reactive superfluid response given by the imaginary component of the complex conductivity, $\sigma_2(\omega) = (\mu_0\omega \lambda^2)^{-1}$ \cite{charpentier2025universal}, where $\mu_0$ is the vacuum magnetic permeability, $\omega$ is taken to be $2\pi f_\mathrm{r, m}$, and $\lambda$ is the magnetic penetration depth, which is $(1.78 \pm 0.02) \ \upmu$m for $\beta$-Ta (Figure 2c, Main Text). Figure \ref{figure: inductive_losses}c presents $Q_\mathrm{int}$ measured at low $P_\mathrm{in}$ and $T$ versus $\sigma_2(\omega)$ for all $\beta$-Ta resonators. The data follow an approximately linear trend with slope $\kappa = (0.16 \pm 0.02) \ \Omega $m, consistent with the ``universal'' $Q_\mathrm{int}^{\mathrm{max}} \approx \kappa \sigma_2$ envelope observed for planar resonators in the high kinetic inductance regime across several superconductors \cite{charpentier2025universal}. 

\section{Resonator transmission measurement} \label{sec: transmission}

All resonators are capacitively coupled to a transmission line in the hanger configuration (Figure \ref{figure: resonator_designs}). The complex transmission coefficient ($S_{\mathrm{21}}$) is measured as a function of probe frequency ($f$) with a VNA over a narrow span around resonance and fit to a model composed of a background term and a resonant response \cite{probst2015efficient}:

\begin{equation} \label{eqn: circle_fit}
    S_\mathrm{21} (f) = \big[Ae^{i(\alpha + 2\pi f \tau)}\big] \   \Bigg[ 1 - \frac{(Q_\mathrm{l}\, / \, |Q_\mathrm{c}|)\ e^{i\phi}}{1 + 2iQ_\mathrm{l}\,(f/f_\mathrm{r, m} - 1)}  \Bigg].
\end{equation}

The background fit parameters are an amplitude $A$, phase $\alpha$, and cable delay $\tau$, and the resonance fit parameters are the loaded quality factor $Q_\mathrm{l}$, the magnitude of the coupling quality factor $|Q_\mathrm{c}|$ and its asymmetry phase $\phi$, and the measured resonance frequency $f_\mathrm{r, m}$. The internal quality factor is calculated using $Q_\mathrm{int}^{-1} = Q_\mathrm{l}^{-1} - |Q_\mathrm{c}|^{-1} \mathrm{cos} \, \phi$. The $S_{\mathrm{21}}$ data are fit to Equation \ref{eqn: circle_fit} using a multi-step fitting procedure specified in Reference \cite{probst2015efficient}. Traces exhibiting signs of nonlinearity are excluded from analysis.

As the resonant response forms a circle in the complex plane, we acquire $S_{\mathrm{21}}(f)$ using a homophasal frequency point distribution (HPD) that samples this circle at uniform phase increments \cite{baity2024circle}. The HPD is generated using

\begin{equation}
    f(\theta) = f_\mathrm{r, m} \bigg( 1 - \frac{1}{2Q_\mathrm{l}} \, \tan \,(\theta/2) \bigg),
\end{equation}

where $\theta \in [0, 2\pi]$ parametrizes the angular position around the resonance circle. We use 251 uniformly spaced $\theta$ points within a span of approximately 15 linewidths ($\Delta f \equiv f_\mathrm{r, m}/Q_\mathrm{l}$) around $f_\mathrm{r, m}$, sampling around $95\%$ of the angular arc of the resonance circle, while excluding far-off-resonant frequencies. We implement the HPD using the VNA's segmented sweep mode, with each point in the $f(\theta)$ array configured as a single-point segment. We use a dwell time of 10 ms per point to allow the source and the receiver to settle at each frequency sweep point.

Generating the HPD requires estimates of $Q_\mathrm{l}$ and $f_\mathrm{r, m}$ at each power and temperature setpoint, as both parameters vary with power and temperature. At each new temperature, we first acquire a coarse $S_{\mathrm{21}}(f)$ trace using a linearly spaced frequency sweep at the highest power within the linear regime, and fit this trace to obtain seed values of $Q_\mathrm{l}$ and $f_\mathrm{r, m}$ to generate the HPD for this power. We then measure a descending sequence of powers, and propagate the $Q_\mathrm{l}$ and $f_\mathrm{r, m}$ extracted at the immediately preceding higher power measurement to generate the HPD for the subsequent lower power measurement. This strategy keeps the HPD centered on the resonance and scaled to its current linewidth, while avoiding the long acquisition times and low signal-to-noise ratio associated with coarse scans at low powers.

\section{Kinetic inductance simulations} \label{sec: simulations}

We obtain the simulated resonance frequency ($f_\mathrm{r, s}$) using full-wave electromagnetic simulations that include only the capacitive and inductive contributions of the resonator geometry. To determine $f_\mathrm{r, s}$, we model the metal as a zero-thickness perfect conductor sheet, compute the input impedance $Z_\mathrm{11}(\omega)$ using a planar method-of-moments solver (Cadence AWR AXIEM), and define $f_\mathrm{r, s}$ as the frequency where $\mathrm{Im}[Z_\mathrm{11}(\omega)] = 0$, where $\omega = 2\pi f$. In these simulations, sapphire is modeled as a uniaxially anisotropic dielectric with in‑plane permittivities $\epsilon_x = \epsilon_y = 9.3$ and out‑of‑plane permittivity $\epsilon_z = 11.5$. 

The measured resonance frequency ($f_\mathrm{r, m}$) of $\beta$-Ta resonators is significantly lower than their $f_\mathrm{r, s}$ (Table \ref{table:resonators}; Figure 2a, Main Text). We quantify this shift to lower frequencies by $\alpha = 1 - (f_\mathrm{r, m} / f_\mathrm{r, s})^2$, which we interpret as the effective kinetic inductance fraction, and hypothesize that the frequency shift originates entirely due to an additional kinetic inductance contribution from the $\beta$-Ta films, which has been reported in other studies \cite{kouwenhoven2023resolving, de2025recombination}. We observe that the CPW resonators exhibit especially high $\alpha$ values, with $\alpha$ increasing with decreasing CPW gap width ($s$) and film thickness ($t$) (Figure 2b, Main Text). We model this behavior by treating the inductance as a series sum of geometric ($L_\mathrm{g}$) and kinetic ($L_\mathrm{k}$) contributions: 

\begin{equation} \label{eqn: alpha}
    \alpha(s, t) = \frac{L_\mathrm{k}(s, t)}{L_\mathrm{k}(s, t) + L_\mathrm{g}(s)}.
\end{equation}

Here, $L_\mathrm{g}$ is calculated as one half of the slope of $\mathrm{Im}[Z_\mathrm{11}(\omega)]$ at $\omega = 2\pi f_\mathrm{r, s}$. Although $L_\mathrm{k}(s, t)$ also varies with the center conductor width ($w$) \cite{gao2008physics}, we control for this dependence by choosing a $w$ for each $s$ that fixes the characteristic impedance to $50 \ \Omega$. $L_\mathrm{k}(s, t)$ is factorized into lateral and vertical components that depend on the resonator geometry and film thickness respectively,

\begin{equation} \label{eqn: lk}
    L_\mathrm{k}(s, t) = G(s) \cdot L_{\mathrm{k/\square}}(t), 
\end{equation}

\noindent where $G(s)$ is a dimensionless geometric factor that maps the resonator's lateral current distribution to the film's sheet kinetic inductance $L_\mathrm{k/\square}(t)$. Equation \ref{eqn: lk} assumes local linear electrodynamics, where $l \ll \xi_0$ (BCS dirty limit), $ql \ll 1$, and $\hbar \omega \ll 2 \Delta$, where $l$ is the electronic mean free path, $\xi_0$ is the BCS coherence length, $q$ is the electromagnetic spatial wavevector of the microwave field, $\hbar$ is the reduced Planck constant, and $\Delta$ is the superconducting gap. These conditions are expected at microwave frequencies for our $\beta$-Ta films; from the measured $\rho_\mathrm{n} = (180 \pm 1) \ \upmu \Omega \ \mathrm{cm}$ (Figure \ref{figure: si_film_characterization}c), a free-electron transport estimate gives $l < 1$ nm. 

\begin{figure}[htbp!]
\includegraphics[width=0.33\textwidth]{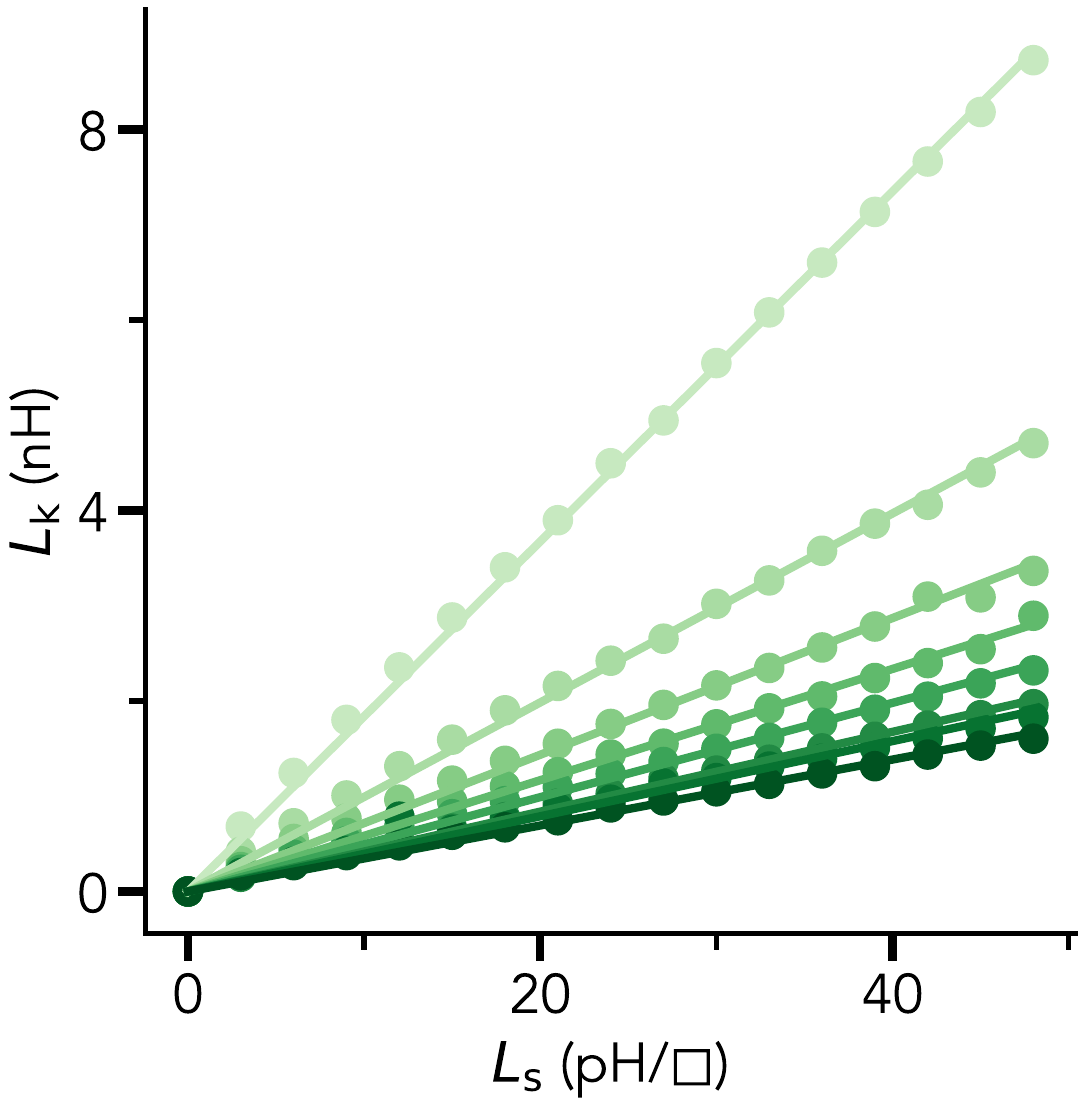}
\caption{\textbf{Extracting the geometric factor $G(s)$ from electromagnetic simulations.} Kinetic inductance $L_\mathrm{k}$ versus applied surface inductance $L_\mathrm{s}$ for 8 CPW geometries. In the simulations, $L_\mathrm{s}$ is applied using a surface impedance boundary $Z_\mathrm{s} = j\omega L_\mathrm{s}$. Colors encode CPW gap width $s$ (light to dark green for $s = 2 \ \upmu$m to $s = 16 \ \upmu$m respectively in steps of $2 \ \upmu$m). Linear fits (solid lines) are used to obtain the slope $G(s)$ for each CPW geometry.}
\label{fig: si_ki_sim}
\end{figure}

To determine $G(s)$ for each resonator geometry, we model the metal with a zero-thickness local surface impedance boundary that is purely inductive, $Z_\mathrm{s} = j\omega L_\mathrm{s}$, and compute the total inductance ($L_\mathrm{tot}$) over a range of $L_\mathrm{s}$ values, where $L_\mathrm{tot}$ is one half of the slope of $\mathrm{Im}[Z_\mathrm{11}(\omega)]$ at resonance. Plotting $L_\mathrm{k} = L_\mathrm{tot} - L_\mathrm{g}$ versus $L_\mathrm{s}$ yields a linear relation (Figure \ref{fig: si_ki_sim}), where the fitted slope is the geometric factor $G(s)$ \cite{gao2008physics}. For our CPW geometries, $G(s)/L$, where $L$ is the CPW center conductor length, is approximately inversely proportional to the transverse dimension $(w + s)$; this relationship allows us to calculate an expected $\alpha$ for each $s$ in Figure 2b of the main text.

The $L_\mathrm{k}(t)$ extracted from each resonator is fit to $L_\mathrm{k/\square}(t) = \mu_0\lambda \coth(t/\lambda)$, where $\mu_0$ is the vacuum magnetic permeability and $\lambda$ is the magnetic penetration depth. We obtain a fitted value of $\lambda = (1.78 \pm 0.02) \ \upmu$m for $\beta$-Ta (Figure 2c, Main Text), which is about an order of magnitude larger than that of $\alpha$-Ta \cite{greytak1964penetration}. We compare the fitted $\lambda$ to the zero-temperature BCS dirty limit estimate given by \cite{watanabe1994kinetic}

\begin{equation}
    \lambda_\mathrm{dirty} = \sqrt{\frac{\hbar \rho_\mathrm{n}}{\pi \mu_0 \Delta_0}},
\end{equation}

\noindent where $\Delta_0$ is the zero-temperature superconducting energy gap. Using the weak-coupling BCS approximation $\Delta_0 \approx 1.76 k_\mathrm{B}T_\mathrm{c}$, where $k_\mathrm{B}$ is the Boltzmann constant, and the measured values of $\rho_\mathrm{n} = (180 \pm 1) \ \upmu \Omega \ \mathrm{cm}$ and $T_\mathrm{c} = (0.7 \pm 0.1)$ K (Figure 1c, Main Text), we find $\lambda_\mathrm{dirty} \approx 1.7 \ \upmu$m, which agrees closely with the fitted $\lambda$. The $0.1 \ \upmu$m difference between $\lambda_\mathrm{dirty}$ and the fitted $\lambda$ likely arises because we overestimate $L_\mathrm{k/\square}$ by ignoring finite-thickness corrections to $G(s)$, particularly in resonators with $s \sim t$ \cite{gao2008physics, clem2013inductances}. 

\section{Resonator microwave loss model} \label{sec: microwave_losses}

We quantify microwave losses in $\beta$-Ta resonators by measuring their internal quality factor ($Q_\mathrm{int}$) at various VNA probe powers ($P_\mathrm{in}$) and mixing chamber stage temperatures ($T$). We convert $P_\mathrm{in}$ to an effective intracavity photon number ($\overline{n}$) by measuring the line attenuation in the dilution refrigerator measurement chain (Figure \ref{fig: wiring_chain}) at room temperature. We model the inverse of $Q_\mathrm{int}(\overline{n}, \ T)$ as a sum of inverse quality factors associated with two-level systems ($Q_\mathrm{TLS}(\overline{n}, \ T)$), thermal quasiparticles ($Q_{\mathrm{QP}}(T)$), and other power- and temperature-independent channels ($Q_\mathrm{other}$) \cite{crowley2023disentangling}:

\begin{equation} \label{eqn: qint}
    \frac{1}{Q_\mathrm{int}} = \frac{1}{Q_\mathrm{TLS}(\overline{n}, T)} + \frac{1}{Q_\mathrm{QP} (T)} + \frac{1}{Q_\mathrm{other}}.
\end{equation}

Here, $Q_\mathrm{TLS}(\overline{n}, \ T)$ is given by

\begin{equation}
    Q_\mathrm{TLS}(\overline{n}, T) = Q_\mathrm{TLS, 0} \frac{\sqrt{1 + \frac{\overline{n}^{\beta_2}}{DT^{\beta_1}} \tanh \big(\frac{\hbar \omega}{2 k_\mathrm{B}T} \big)}}{\tanh \big(\frac{\hbar \omega}{2 k_\mathrm{B}T} \big)},
\end{equation}

\noindent where $Q_\mathrm{TLS, 0}$ is the unsaturated TLS quality factor that quantifies the inverse linear absorption due to TLSs, $D$, $\beta_1$, and $\beta_2$ are fit parameters characterizing TLS saturation, $\omega = 2\pi f_\mathrm{r, m}$ is the resonator's angular frequency, and $k_\mathrm{B}$ is the Boltzmann constant; and $Q_\mathrm{QP} (T)$ is

\begin{equation}
    Q_\mathrm{QP}(T) = A_\mathrm{QP} \frac{e^{\Delta_0/k_\mathrm{B}T}}{\sinh{\big(\frac{\hbar \omega}{2 k_\mathrm{B}T} \big)} K_\mathrm{0}\big(\frac{\hbar \omega}{2 k_\mathrm{B}T} \big)},
\end{equation}

\noindent where $A_\mathrm{QP}$ is a device-specific, temperature-independent prefactor, $\Delta_0$ is the zero-temperature superconducting energy gap, and  $K_\mathrm{0}$ is the zeroth-order modified Bessel function of the second kind. We fit the full dataset $Q_\mathrm{int}(\overline{n}, \ T)$ using a self-consistent routine to extract well-defined $Q_\mathrm{TLS, 0}$ and $Q_\mathrm{other}$ values that show weak correlations with other parameters; these are reported in Table \ref{table:resonators}.

\section{Surface TLS loss in \texorpdfstring{$\beta$-Ta}{beta-Ta} resonators} \label{sec: surface_tls_loss}

We quantify the relative contributions of  surface and bulk TLS baths by studying the dependence of the unsaturated TLS quality factor $Q_\mathrm{TLS, 0}$ on the electric field surface participation ratio (SPR) \cite{wang2015surface}. For CPW resonators, we vary the SPR by changing the gap width $s$, and for LE resonators, by changing the area of and spacing between capacitor pads. The metal-air (MA), metal-substrate (MS), and substrate-air (SA) interfaces of the resonator are modeled as 3 nm-thick dielectric layers with relative permittivity  $\epsilon_\mathrm{r} = 10$. Using finite element simulations (Ansys Maxwell), we compute the electric field energy stored in these layers to determine the interface-specific participation ratios $p_\mathrm{MA}$, $p_\mathrm{MS}$, and $p_\mathrm{SA}$, following methods detailed in Reference \cite{crowley2023disentangling}.  

\begin{figure}[htbp!]
\includegraphics[width=0.67\textwidth]{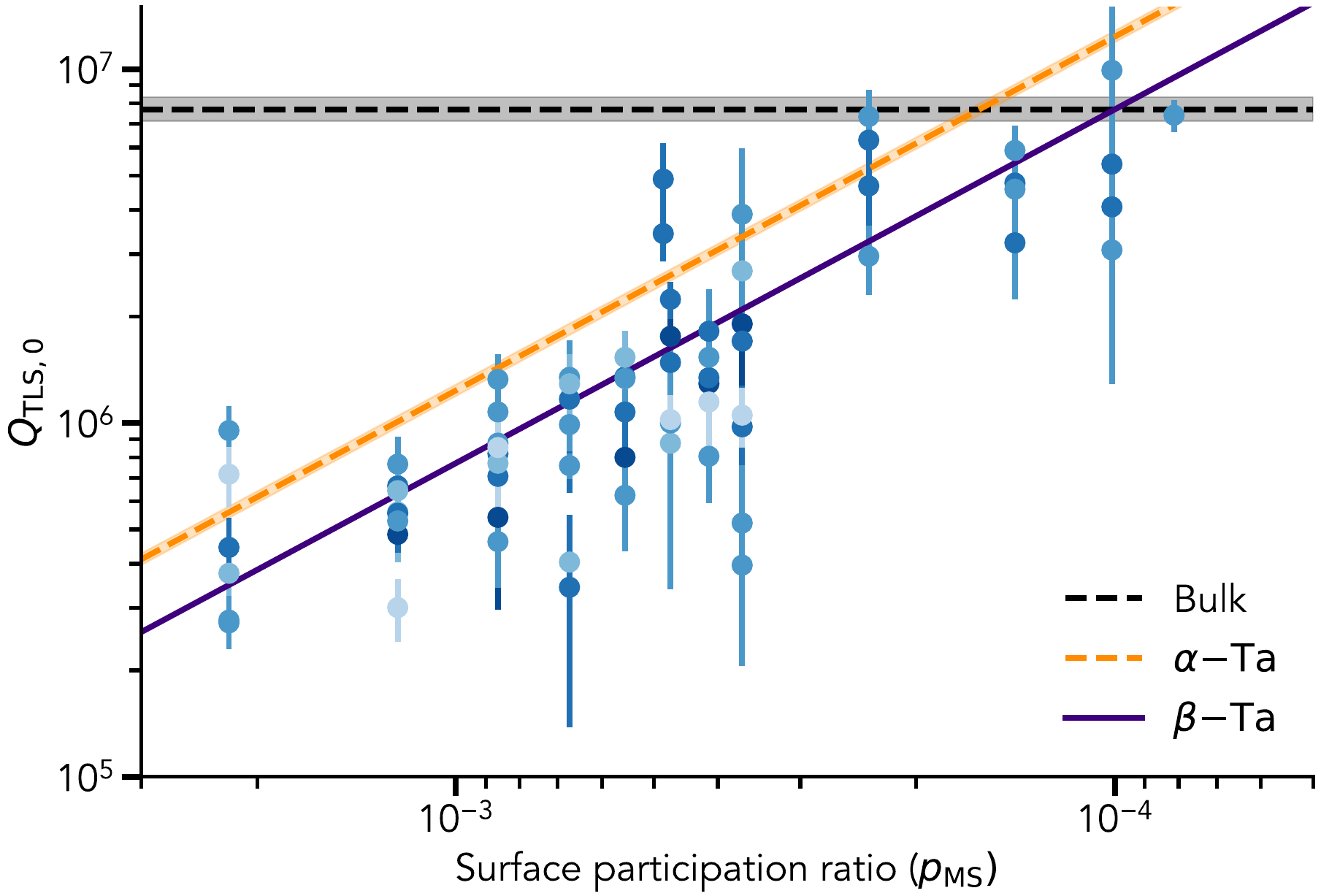}
\caption{\textbf{Surface TLS loss in $\beta$-Ta resonators}. $Q_\mathrm{TLS, 0}$ as a function of $p_\mathrm{MS}$ for 71 resonators, yielding a single $\tan{\delta}$ (solid purple line), regardless of metal film thickness. Film thickness ($t$) is encoded by shades of blue (lightest blue: $t = 0.02 \ \upmu$m, darkest blue: $t = 2.00 \ \upmu$m). Error bars of $Q_\mathrm{TLS, 0}$ represent the s.e. estimated from the covariance matrix of the fit to Equation \ref{eqn: qint}. The loss tangents corresponding to the $\alpha$-Ta surface (dashed orange line) and the bulk substrate (dashed black line) and their uncertainties (shaded regions) are added from Reference \cite{crowley2023disentangling} for comparison.}
\label{fig: si_tls_losses_thickness}
\end{figure}

We find that the $Q_\mathrm{TLS, 0}$ of 71 resonators scales inversely with $p_\mathrm{MS}$, with the full dataset being well-described by a single effective surface loss tangent ($\tan \delta$), regardless of metal film thickness (Figure \ref{fig: si_tls_losses_thickness}). The data is fit to \cite{crowley2023disentangling}

\begin{equation}
\begin{split}
    Q_\mathrm{TLS, 0}^{-1} &= p_\mathrm{MS} \tan{\delta_\mathrm{MS}} + p_\mathrm{MA} \tan{\delta_\mathrm{MA}} + p_\mathrm{SA} \tan{\delta_\mathrm{SA}} \\
    &= p_\mathrm{MS} (\tan{\delta_\mathrm{MS}} + (p_\mathrm{MA}/p_\mathrm{MS})\tan{\delta_\mathrm{MA}} + (p_\mathrm{SA}/p_\mathrm{MS})\tan{\delta_\mathrm{SA}}) 
    \\
    &= p_\mathrm{MS} \tan{\delta}.
\end{split}
\end{equation}

\begin{figure}[htbp!]
\includegraphics[width=\textwidth]{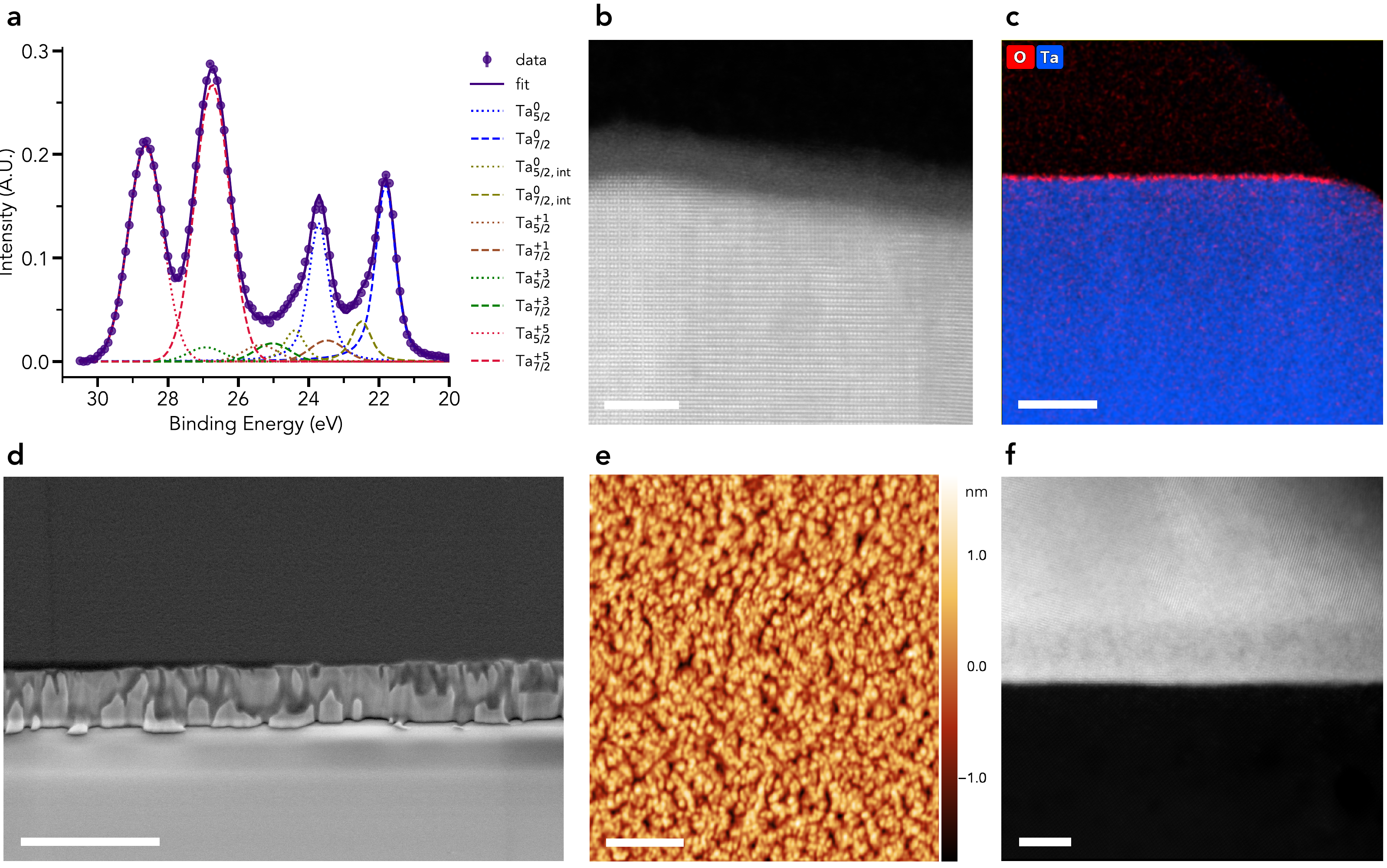}
\caption{\textbf{Characterization of the MA and MS interfaces in $\beta$-Ta resonators.} (a) Ta4f XPS spectrum fitted to a multi-component model (solid line). Dashed and dotted lines represent fits to the individual $\mathrm{Ta4f}_{7/2}$ and $\mathrm{Ta4f}_{5/2}$ peaks respectively. We identify the following species, with the brackets reporting the fitted binding energies in eV of the $\mathrm{Ta4f}_{7/2}$ and $\mathrm{Ta4f}_{5/2}$ peaks respectively: $\mathrm{Ta}^0$ (21.8, 23.7), $\mathrm{Ta}^{1+}$ (23.5, 25.4), $\mathrm{Ta}^{3+}$ (25.0, 26.9), $\mathrm{Ta}^{5+}$ (26.7, 28.6), and $\mathrm{Ta}^{0}_\mathrm{int}$ (22.5, 24.4). (b) Cross-sectional STEM image of the MA interface of a $\beta$-Ta film shows a roughly 3 nm-thick native oxide layer (darker gray) over the Ta atoms (lighter gray). Scale bar: 5 nm. (c) False-colored EDS elemental map confirming the presence of the native oxide layer (red) over the Ta film (blue). Scale bar: 50 nm. (d) SEM image of the etched sidewall of a $\beta$-Ta resonator showing rough protrusions with lateral dimensions of about 100 nm. The SEM image is acquired in the secondary electron mode using an accelerating voltage of 5 kV, probe current of 25 pA, magnification of 50000X, working distance of 4.6 mm, and a dwell time of 10 $\upmu$s. Scale bar: 1 $\upmu$m. (e) False-colored AFM scan of the surface of a $\beta$-Ta film, revealing grains with a size of approximately 10 nm. Scale bar: 100 nm. (f) Cross-sectional STEM image of the MS interface between $\beta$-Ta (lighter gray) and sapphire (black) suggesting a roughly 5 nm-thick amorphous interfacial layer (darker gray). Scale bar: 5 nm.}
\label{fig: si_ma_ms_interface}
\end{figure}

We extract a $\tan{\delta} = (1.6 \pm 0.1) \times 10^{-3}$, which implies that the surface TLS loss in $\beta$-Ta is about twice that reported for $\alpha$-Ta, ($8.1 \pm 0.6) \times 10^{-4}$ \cite{crowley2023disentangling}, under the same surface processing conditions of a post-fabrication piranha clean followed by a buffered oxide etch treatment (Section \ref{sec: device fab}). We perform additional surface and interface characterization to uncover possible explanations for the higher apparent surface TLS loss in $\beta$-Ta (Figure \ref{fig: si_ma_ms_interface}).

We use X-ray photoelectron spectroscopy (XPS) to determine the composition of the native tantalum oxide ($\mathrm{TaOx}$) at the MA interface. We measure Ta4f core‑level spectra (Figure \ref{fig: si_ma_ms_interface}a) using an X-ray photoelectron spectrometer (Thermo Scientific K-Alpha) equipped with an Al K$\alpha = 1486$ eV X-ray source. The spectra are averaged over 20 individual scans, each acquired with a spot size of 400 $\upmu$m, a step size of 0.1 eV, and a dwell time per point of 100 ms. We subtract a Shirley background over $20-30$ eV and then fit the data to a multi-component model containing the $\mathrm{Ta4f}_{5/2}$ and $\mathrm{Ta4f}_{7/2}$ doublets for metallic $\mathrm{Ta}^0$, oxide species $\mathrm{Ta}^{1+}$, $\mathrm{Ta}^{3+}$, and $\mathrm{Ta}^{5+}$, as well as an interface species $\mathrm{Ta}^0_{\mathrm{int}}$ \cite{mclellan2023chemical}. We determine the relative peak intensities for each component as the ratio of the area under each fitted peak doublet to the total area under all fitted peak doublets. Finally, we estimate the effective oxide thickness $t_\mathrm{TaOx}$ using the standard expression for overlayer films on a substrate \cite{jablonski2020effective}, 

\begin{equation}
    t_\mathrm{TaOx} = \lambda_\mathrm{eff} \ln \bigg( \frac{N_\mathrm{m}}{N_\mathrm{ox}} \frac{I_\mathrm{ox}}{I_\mathrm{m}} + 1 \bigg), 
\end{equation}

\noindent where $\lambda_\mathrm{eff}$ is the effective attenuation length of Ta4f photoelectrons, $I_\mathrm{ox}$ is the sum of the relative peak intensities of the $\mathrm{Ta}^{1+}$, $\mathrm{Ta}^{3+}$, and $\mathrm{Ta}^{5+}$ oxide species, $I_\mathrm{m}$ is the relative peak intensity of the metallic $\mathrm{Ta}^{0}$ species, and $N_\mathrm{m}$ and $N_\mathrm{ox}$ are the volume densities of the Ta atoms in the metal and oxide layers respectively. Here, the photoelectron emission angle is perpendicular to the film surface. We calculate $\lambda_\mathrm{eff} = 1.51$ nm \cite{jablonski2020effective}, and use $N_\mathrm{m} = 5.43 \times 10^{22} \ \mathrm{cm}^{-3}$ and $N_\mathrm{ox} = 2.24 \times 10^{22} \ \mathrm{cm}^{-3}$. We find $t_\mathrm{TaOx} = (2.9 \pm 0.1)$ nm, which is consistent with cross-sectional STEM images (Figure \ref{fig: si_ma_ms_interface}b) and energy-dispersive X-ray spectroscopy of the MA interface (Figure \ref{fig: si_ma_ms_interface}c).

The $t_\mathrm{TaOx}$ obtained for $\beta$-Ta is about 1.2 times larger than that reported for $\alpha$-Ta, $(2.45 \pm 0.07)$ nm \cite{mclellan2023chemical}, under the same post-fabrication surface processing conditions. Treating the MA interface loss as arising from the native $\mathrm{TaOx}$ overlayer and scaling linearly with $t_\mathrm{TaOx}$, and assuming the same rescaled intrinsic loss tangent for $\mathrm{TaOx}$ across $\alpha$-Ta and $\beta$-Ta, $(p_\mathrm{MA}/p_\mathrm{MS}) \tan \delta_\mathrm{TaOx} = (6 \pm 1) \times 10^{-4}$ \cite{crowley2023disentangling}, the increased $t_\mathrm{TaOx}$ accounts for about $20\%$ of the higher apparent surface loss observed in $\beta$-Ta resonators.

We identify three plausible contributors for the remaining excess surface loss. Scanning electron microscopy (SEM; Thermo Scientific Verios 460 XHR) of etched device sidewalls shows rough protrusions with lateral dimensions of about $100$ nm (Figure \ref{fig: si_ma_ms_interface}d) and atomic force microscopy (AFM; Bruker Dimension Icon) of the film surface reveals grain sizes of approximately $10$ nm (Figure \ref{fig: si_ma_ms_interface}e). Additional oxide may form over the protrusions and along the grain boundaries, which increases the effective oxide volume at the MA interface, and can lead to enhanced surface losses, especially in regions with high electric field participation. In comparison, the average grain size in $\langle110\rangle$-oriented $\alpha$-Ta films is larger at about $50-200$ nm \cite{bahrami2026vortex}. Additionally, cross-sectional STEM images of the MS interface show an approximately 5 nm-thick amorphous layer at the interface between $\beta$-Ta and sapphire (Figure \ref{fig: si_ma_ms_interface}f), whereas the epitaxial interface between $\alpha$-Ta and sapphire is lattice-matched \cite{place2021new}. As $p_\mathrm{MA}$ and $p_\mathrm{MS}$ do not vary independently across our resonator geometries, we are unable to quantify the relative contributions of these features to the observed surface losses. 

\section{\texorpdfstring{$\beta$-Ta}{beta-Ta} on silicon}

\begin{figure}[htbp!]
\includegraphics[width=\textwidth]{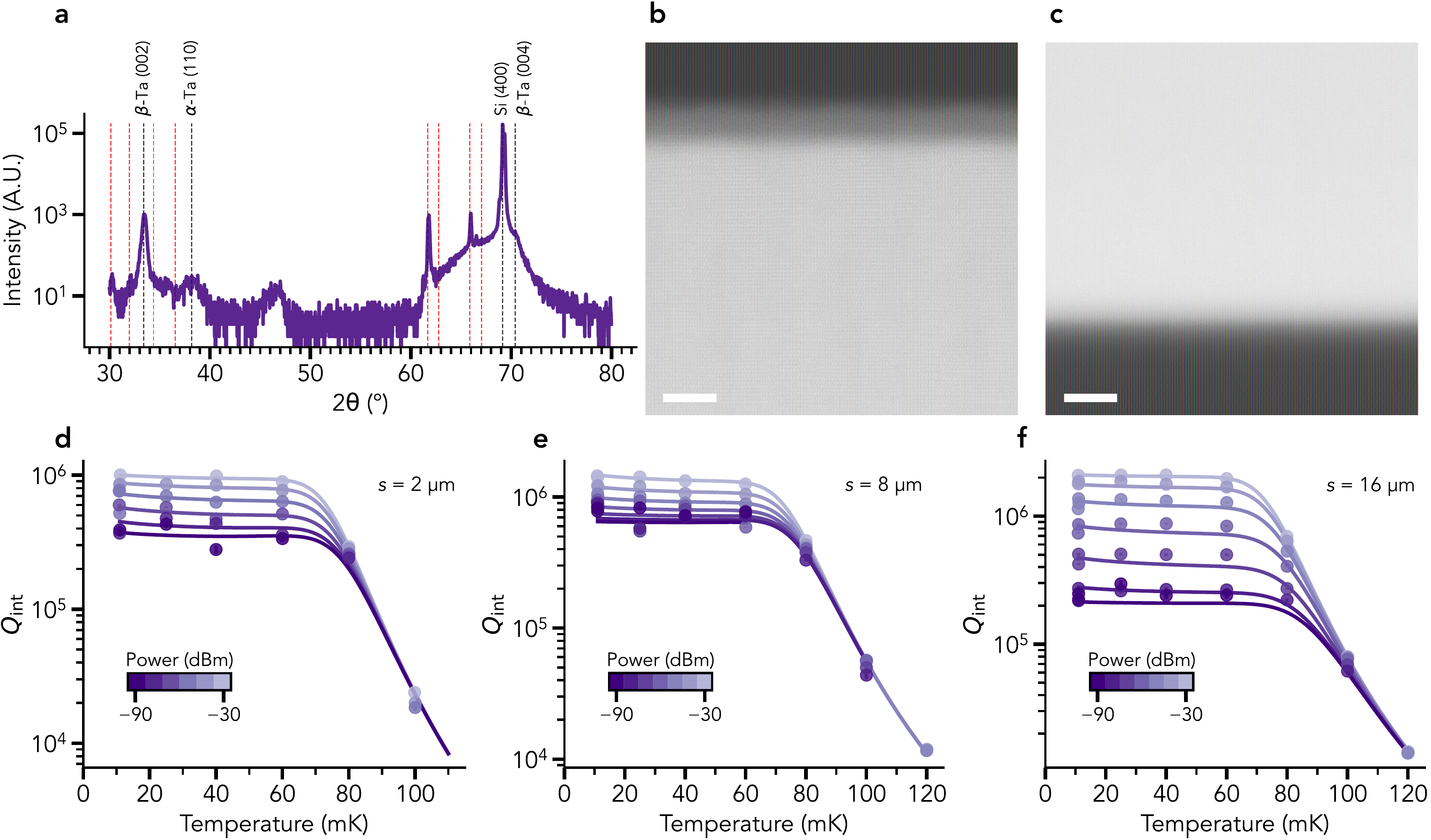}
\caption{\textbf{$\beta$-Ta-on-Si resonators.} (a) XRD pattern obtained from a Ta film deposited on high-purity silicon. Dashed black vertical lines indicate the following peak locations: $\beta$-Ta $(002)$ at $33.4 \degree$, $\alpha$-Ta $(110)$ at $38.2 \degree$, $\mathrm{Si} \ (400)$ at $69.2 \degree$, and $\beta$-Ta $(004)$ at $70.4 \degree$. The spectrum reveals $\beta$-Ta as the predominant phase, with traces of $\alpha$-Ta. The weak, broad peak at $46.5 \degree$ is absent in an XRD pattern obtained from the bare Si substrate, and does not match any reported peaks for $\beta$-Ta. Dashed red vertical lines indicate artifact peaks arising from Cu $K_\upbeta$ leakage and tungsten contamination. Cross-sectional STEM images of the (b) MA interface shows a roughly 2.5 nm-thick native oxide layer (darker gray) over the Ta atoms (lighter gray). Scale bar: 5 nm. (c) MS interface between $\beta$-Ta (light gray) and Si (black), suggesting the presence of a roughly 1 nm-thick amorphous Ta layer at the epitaxial interface. Scale bar: 5 nm. Power- and temperature-dependence of $Q_\mathrm{int}$ with fits (solid lines) to the loss model described in Section \ref{sec: microwave_losses} for CPW resonators with (d) $s = 2 \ \upmu$m (F17--CPW--2 in Table \ref{table:beta_ta_si_resonators}), (d) $s = 8 \ \upmu$m (F17--CPW--8), and (f) $s = 16 \ \upmu$m (F17--CPW--16).}
\label{fig: beta_ta_on_si}
\end{figure}

Recent work demonstrates that transmon qubits made from $\alpha$-Ta films deposited on high-purity silicon (Si) have relaxation and coherence times exceeding one millisecond, enabled by the lower apparent bulk dielectric loss of high-purity Si compared to that of sapphire \cite{bland2025millisecond}. A natural question is whether the performance of $\beta$-Ta-based transmon qubits can be similarly improved by using high-purity Si instead of sapphire. We sputter a 200 nm-thick Ta film (Film F17) on $1'' \times 1''$ pieces of intrinsic float zone Si wafers (Siegert Wafer GmbH) with nominal resistivity $>20 \ \mathrm{k}\Omega$ cm  using the same room-temperature deposition procedure described in Section \ref{sec: film dep}, except for two changes. First, we remove the native Si surface oxide after piranha-cleaning the substrate and before loading it into the sputtering system, using the following recipe: the substrate is immersed in $5\%$ hydrofluoric acid for 3 min, then rinsed three times in separate beakers of deionized water, and finally blow dried under nitrogen flow. Second, the substrate is heated in situ at 600 $\degree$C, instead of 300 $\degree$C, for 1 hour and then allowed to cool to room temperature prior to deposition. We verify the crystallographic phase of the film using XRD (Figure \ref{fig: beta_ta_on_si}a) and obtain STEM images of the MA (Figure \ref{fig: beta_ta_on_si}b) and MS (Figure \ref{fig: beta_ta_on_si}c) interfaces. We measure 7 CPW resonators with gap widths $s = 2-16 \ \upmu$m made from this film and confirm that their kinetic inductance is consistent with the values expected from Figure 2 of the main text.

Microwave loss measurements (Section \ref{sec: microwave_losses}) reveal power- and temperature-independent losses ($Q_\mathrm{other}^{-1}$) that limit the $Q_\mathrm{int}$ at high powers and millikelvin temperatures to a range of $0.8-2.8$ million across all resonators, independent of $p_\mathrm{MS}$ (Figures \ref{fig: beta_ta_on_si}d--f). Most resonators show power-dependent losses at millikelvin temperatures, with $Q_\mathrm{TLS, 0}$ increasing from $(0.37 \pm 0.24) \times 10^6$ for $s = 2 \ \upmu$m (Figure \ref{fig: beta_ta_on_si}d) to $(0.63 \pm 0.22) \times 10^6$ for $s = 8 \ \upmu$m (Figure \ref{fig: beta_ta_on_si}e). However, resonators with $s = 10 \ \upmu$m and $s = 12 \ \upmu$m show weak power dependence due to a low $Q_\mathrm{other}$ (Table \ref{table:beta_ta_si_resonators}), resulting in unreliable fitted values of $Q_\mathrm{TLS, 0}$, while the resonator with $s = 16 \ \upmu$m shows anomalously low $Q_\mathrm{TLS, 0}$ (Figure \ref{fig: beta_ta_on_si}f). The significantly lower $Q_\mathrm{other}$ observed in $\beta$-Ta resonators on Si versus sapphire motivates further careful investigation of the MS interface between $\beta$-Ta and Si to identify the precise nature of this loss. Despite the low $Q_\mathrm{other}$, our resonators record the highest $Q_\mathrm{int}$ across $\beta$-Ta-on-Si resonators reported in the literature \cite{singer2024tantalum, urade2024microwave, arlt2025high}.

\begin{table}[H]
    \centering
    \caption{\textbf{Summary of microwave loss measurements on $\beta$-Ta-on-Si resonators.} Resonators are named according to the convention used in Table \ref{table:resonators}. $p_\mathrm{MS}$ is computed using an $\epsilon_\mathrm{r} = 11.68$ for Si and a Si trench depth of 100 nm measured by SEM imaging, following methods described in Reference \cite{bland2025millisecond}. $Q_\mathrm{TLS, 0}$ and $Q_\mathrm{other}$ are omitted where fit errors exceeded the fitted values. Measurements on the resonator with $s = 14 \ \upmu$m are omitted as they showed unphysical $Q_\mathrm{int}$ values because the resonator was extremely overcoupled to the transmission line.}
    \label{table:beta_ta_si_resonators}
    \begin{tabular*}{\columnwidth}
    {l@{\extracolsep{\fill}} 
    c@{\extracolsep{\fill}}
    c@{\extracolsep{\fill}}
    c@{\extracolsep{\fill}}
    c@{\extracolsep{\fill}}
    c@{\extracolsep{\fill}}
    c@{\extracolsep{\fill}}c}
        \toprule
        Resonator & $p_\mathrm{MS} \ (\times10^{-4})$ & $|Q_\mathrm{c}| \ (\times10^6)$ & $f_\mathrm{r, m}$ (GHz) & $Q_\mathrm{TLS, 0} \ (\times10^6)$ & $Q_\mathrm{other} \ (\times10^6)$ \\
        \midrule
        F17--CPW--2 & 18.8 & 4.78 & 4.35 & $0.37 \pm 0.24$ & $1.33 \pm 0.23$ \\
        F17--CPW--4 & 10.9 & 0.99 & 4.74 & $0.77 \pm 0.59$ & $2.76 \pm 1.98$  \\
        F17--CPW--6 & 7.89 & 0.52 & 5.21 & $0.89 \pm 0.70$ & $2.67 \pm 0.70$ \\
        F17--CPW--8 & 6.23 & 0.38 & 5.64 & $0.63 \pm 0.22$ & ---  \\
        F17--CPW--10 & 5.19 & 0.23 & 6.14 & --- & $1.40 \pm 0.06$ \\
        F17--CPW--12 & 4.47 & 0.19 & 6.44 & --- & $0.84 \pm 0.03$ \\
        F17--CPW--16 & 3.52 & 0.086 & 6.91 & $0.21 \pm 0.05$ & $2.49 \pm 0.04$\\
        \bottomrule
    \end{tabular*}
\end{table}

\raggedright

\bibliographystyle{MSP}
\bibliography{references.bib}